\documentclass[smallextended]{svjour3}
\usepackage{natbib,rotating,graphicx}
\DeclareGraphicsExtensions{.eps,.ps,.eps.gz,.ps.gz,.eps.Z}
\begin{document}

\newcommand{\Msun} {$M_{\odot}$}

\title{Liverpool Telescope 2: a new robotic facility for rapid transient follow-up}

\author{C.M.~Copperwheat \and I.A.~Steele \and R.M.~Barnsley \and S.D.~Bates \and D.~Bersier \and M.F.~Bode \and D.~Carter \and N.R.~Clay \and C.A.~Collins \and M.J.~Darnley \and C.J.~Davis \and C.M.~Gutierrez \and D.J.~Harman \and P.A.~James \and J.H.~Knapen \and S.~Kobayashi \and J.M.~Marchant \and P.A.~Mazzali \and C.J.~Mottram \and C.G.~Mundell \and A.~Newsam \and A.~Oscoz \and E.~Palle \and A.~Piascik \and R.~Rebolo \and R.J.~Smith\\\\
}		

\institute{C.M.~Copperwheat \and  I.A.~Steele \and R.M.~Barnsley \and S.D.~Bates \and D.~Bersier \and M.F.~Bode \and D.~Carter \and N.R.~Clay \and C.A.~Collins \and M.J.~Darnley \and C.J.~Davis \and D.J.~Harman \and P.A.~James\and S.~Kobayashi \and J.M.~Marchant \and P.A.~Mazzali \and C.J.~Mottram \and C.G.~Mundell \and A.~Newsam \and A.~Piascik \and R.J.~Smith
		\at Astrophysics Research Institute, Liverpool John Moores University, IC2, Liverpool Science Park, 146 Brownlow Hill, Liverpool, L3 5RF, UK \\\email{c.m.copperwheat@ljmu.ac.uk}
	\and C.M.~Gutierrez \and J.H.~Knapen  \and A.~Oscoz \and E.~Palle \and R.~Rebolo
		\at Instituto de Astrof\'\i sica de Canarias, E-38200 La Laguna,
Tenerife, Spain\\ Departamento de Astrof\'\i sica, Universidad de La Laguna, E-38205 La
Laguna, Tenerife, Spain
		}

\titlerunning{Liverpool Telescope 2}
\authorrunning{Copperwheat et al.}

\date{Received: date / Accepted: date}

\maketitle

\begin{abstract} 
The Liverpool Telescope is one of the world's premier facilities for time domain
astronomy. The time domain landscape is set to radically change in the coming
decade, with synoptic all-sky surveys such as LSST providing huge numbers of
transient detections on a nightly basis; transient detections across the electromagnetic spectrum from other major facilities such as SVOM, 
 SKA and CTA; and
the era of `multi-messenger astronomy', wherein astrophysical events are
detected via non-electromagnetic means, such as neutrino or gravitational wave
emission. We describe here our plans for the Liverpool Telescope 2: a new robotic
telescope designed to capitalise on this new era of time domain astronomy.
LT2 will be a $4$-metre class facility co-located with the
Liverpool Telescope at the Observatorio del Roque de Los Muchachos on the Canary
island of La Palma. The telescope will be designed for extremely rapid response:
the aim is that the telescope will take data within $30$ seconds of the
receipt of a trigger from another facility. The motivation for this is twofold:
firstly it will make it a world-leading
facility for the study of fast fading transients and explosive phenomena
discovered at early times. Secondly, it will enable large-scale programmes of low-to-intermediate resolution spectral classification of transients to be performed with great efficiency. In the target-rich environment of the LSST era, minimising
acquisition overheads will be key to maximising the science gains from any
follow-up programme. The telescope will have a diverse instrument suite
which is simultaneously mounted for automatic changes, but it is envisaged that
the primary instrument will be an intermediate resolution, optical/infrared
spectrograph for scientific exploitation of transients discovered with the next
generation of synoptic survey facilities. In this paper we outline the core
science drivers for the telescope, and the requirements for the optical and
mechanical design. 
\end{abstract}

\section{Introduction}  
The Liverpool Telescope (LT; \citealt{Steele04}) is a $2.0$ metre robotic telescope owned and operated by the Astrophysics Research Institute (ARI) at Liverpool John Moores University (LJMU). It is located at the Observatorio del Roque de Los Muchachos (ORM), on the Canary island of La Palma, Spain. LT's operation is supported by funding from LJMU, the UK Science and Technology Facilities Council, and several other partners. The telescope is managed remotely from the ARI in Liverpool but is completely autonomous throughout the night; the robotic control software is responsible for safe operation of the telescope and chooses what and when to observe from a user-defined list of targets. Science operations began in $2004$, and the telescope is now in a mature phase of operation with a diverse and stable instrument suite\footnote{http://telescope.livjm.ac.uk/}. It is expected the telescope will remain scientifically competitive until at least the beginning of the next decade.

Robotic telescopes are a particularly powerful tool for probing the time domain. Variable objects can be monitored on timescales from seconds to years with minimal user intervention, and the task of coordinating simultaneous  observations with other ground and space based facilities is greatly simplified. Another key advantage of limiting human involvement is a very rapid reaction to unpredictable transient phenomena. This has proven to be a particular strength of the LT: as well as being robotic, the telescope is fast-slewing and features a novel clamshell design enclosure, which means on average the telescope can take data within $180$ seconds of the receipt of a transient alert. The commencement of LT science operations in 2004 coincided somewhat serendipitously with the launch of the {\it Swift} Gamma Ray Burst (GRB) Mission \citep{Gehrels04}. The rapidly-fading nature of GRB counterparts means that rapid reaction is essential for follow-up, and this is an area in which the LT has been particularly productive, contributing to many high-impact papers (e.g. \citealt{Mundell07, Racusin08, Steele09, Abdo10, Thorne11, Mundell13}). The operational life of the LT has also coincided with a rapid advancement in supernova (SN) studies. Wide field synoptic surveys such as the Palomar Transient Factory (PTF; \citealt{Rau09}) have enabled increasingly earlier discovery of large numbers of SNe. The LT has provided early-time photometric and spectroscopic follow-up of many such events (e.g. \citealt{Pastorello07,Mazzali08,Valenti09,Perets10,Nugent11}).

Wide field synoptic surveys have been extremely productive for transient science. The next decade will see upgrades to existing surveys as well as construction of the next generation of facilities. The most ambitious is the Large Synoptic Survey Telescope (LSST: \citealt{Ivezic08}), which is expected to see scientific first-light in 2021. LSST will image the entire Southern sky every three nights on average, and will release of order $10^6$ alerts per night. This new era will require a new approach to follow-up: there will be huge numbers of supernovae discovered at early times and close to statistically complete samples of rare subtypes. At the same time, many of the photometric monitoring observations currently provided by follow-up facilities (such as the LT) will be provided `for free' by the surveys themselves, however follow-up facilities will still be vital for spectroscopic classification and higher cadence photometric monitoring.

The profile of time domain astronomy will be further enhanced in the coming decade with a slew of new missions which will probe temporal variability and detect transient phenomena across the electromagnetic spectrum. These include major new facilities such as the Square Kilometre Array (SKA; \citealt{Carilli04}) and the Cherenkov Telescope Array (CTA: \citealt{Actis11}). Proposed space based missions such as the {\it Swift} successor SVOM \citep{Paul11} will detect high energy transients and provide triggers for rapid ground based follow-up. The next decade will also see the beginning of the era of `multi-messenger' astronomy, in which astrophysical events are detected by non-electromagnetic means, such as via their neutrino emission using the IceCube detector \citep{Karle03}. Advanced LIGO \citep{Harry10} will begin science operations in 2015, followed by Advanced Virgo \citep{Degallaix13} soon after. The array will reach full sensitivity in 2020, and is anticipated to make the first detections of gravitational wave emission. Detection of optical counterparts to gravitational wave events will be key for both verification and scientific exploitation of LIGO and Virgo detections.

This new era of time domain astronomy will require the next generation of robotic telescope for efficient follow-up. Therefore, in Autumn 2012, we began a feasibility study for a new facility, the `Liverpool Telescope 2' (LT2), to come into operation around the beginning of the next decade. In this paper we provide an overall summary of the science drivers for this new telescope, and the telescope design parameters which have been chosen to meet these science requirements. 

\section{Main science drivers}  
\label{sec:science}

\subsection{Supernovae}

\subsubsection{Overview}
\label{sec:snoverview}

Supernovae are of intrinsic interest as spectacular and catastrophic events
which mark the end of the life of the star. They play a major role in the
history of the Universe as end points of the evolution of different types of
stars, as nuclear furnaces where heavy elements are produced and ejected into
the interstellar medium, and as cosmographical beacons. The
most well-known contemporary application is the use of SNe of Type Ia as
standardizable candles to measure cosmic distances. Thanks to a relation between
the intrinsic luminosity of SNe Ia and the width of their light curve, which is
a directly observable quantity \citep{Phillips93,Phillips99}, SNe Ia have been
instrumental in the discovery of dark energy \citep{Riess98,Perlmutter99}.

\subsubsection{Supernovae Ia}

Despite their importance, we still know embarrassingly little about SNe
Ia. The nature of the progenitor system is hotly
debated, spurred by searches for surviving companions \citep{RuizLapuente04},
traces of hydrogen from the companion \citep{Mattila05}, or bumps in the light
curves caused by interaction of the SN with the companion \citep{Kasen04}, none of
which seem to yield conclusive results. The classical single-degenerate (SD) scenario is now being
challenged by the double-degenerate (DD) scenario (see, e.g. \citealt{Schaefer12}); which was previously dismissed as it was thought
to lead simply to an accretion-induced collapse of the white dwarf into a neutron star. The consistent light curves of SN Ia imply the masses of the
progenitors are similar, and very close to the Chandrasekhar mass
\citep{Mazzali07}. However, in recent years a number of SNe have been observed
in which the progenitor mass has been inferred to be significantly greater than
the Chandrasekhar mass (e.g., SN2003fg, \citealt{Astier06,Howell06,Scalzo10}). The early-time spectra of these `super-Chandrasekhar' SNe resemble those of
other SNe Ia while their luminosity is about twice the average. This is difficult to reconcile with the SD paradigm, although attempts have been made (for example, a `supermassive' white dwarf
could be stabilised by rotation: see \citealt{Hachisu12} and references therein). It is quite possible at this stage that there are multiple routes to a SN Ia
explosion, which could have significant consequences for the use of SNe Ia as standard candles. For recent discussions of SN Ia formation models see, e.g., \citet{Wang12,Kushnir13,Maoz13}.

Another key aspect which is not currently understood are the details of the explosion and the
ensuing nucleosynthesis (e.g. the transition from deflagration to detonation). Turbulent deflagration models \citep{Hillebrandt00,Ropke07} of SNe Ia predict a large amount of unburned carbon and oxygen in
the outer layers of the ejecta whereas delayed detonation models
\citep{Hoflich02,Kasen09} predict that the carbon should be burned more
completely and little or no C and O should be left at low velocities, which seems
more in line with observations \citep{Mazzali07}. The optimum time to probe
for the carbon features is one week before maximum light \citep{Parrent11}. SN2011fe was
caught by PTF within hours of the explosion \citep{Nugent11}, and the detection of unburned carbon and high velocity oxygen in the earliest spectrum (obtained with the LT) was the first decisive
evidence that CO white dwarfs are SN Ia progenitors. Other high velocity
features observed in early-time SN Ia spectra include Ca~II and Si~II.
\citep{Mazzali05}. These features imply density enhancements due to a thick disc
or high density companion wind surrounding the exploding white dwarf, which
supports the SD models.

\subsubsection{Core collapse supernovae}

The population of core collapse SNe consists of a wide range of subclasses based on observed
differences in the light curves or spectra of the aftermath of the explosion,
which mostly reflect the size and degree of stripping of the progenitor at the
time of explosion.  It is important that we link these observed differences with
differences in the physical parameters of the progenitors. One of the main
heterogeneities is the presence or absence of hydrogen in the spectra. Type
Ib/c SNe do not show hydrogen lines, (or strong silicon absorption, which
distinguishes them from Type Ia), and this is attributed to the progenitor
losing most of its hydrogen envelope prior to the core collapse. However, it now
appears likely that a fraction of the envelope is lost in some of the
subclasses of Type II SNe as well, with only those of Type IIP (the most common
type; which show a plateau phase in their light curves) retaining most or all of
their envelopes up until the explosion. There are likely many other parameters
which affect the fate of the massive progenitor, such as the initial mass,
metallicity, rotation rate, and the presence and properties of binary companions
or magnetic fields. While there are many uncertainties, the study of core
collapse SNe is more advanced than that of Type Ia SNe in one key aspect: stellar
progenitors have been unambiguously discovered and identified (see \citealt{Smartt09} and references therein). However, this has only been possible for the nearest SNe, and so has not
provided any useful constraints on less common SN types. Also, no
progenitor of any SNe Ic have been detected, despite their importance as hosts of
GRBs in some cases, probably because these stars are intrinsically blue \citep{Smartt09}. Studies of SNe environments have also been productive:
\citet{Anderson08} showed that the positions of SNe Ib/c in late-type galaxies
correlate with H$\alpha$ emission, but this is not true of Type II SNe.
Since this emission requires a young population of massive stars, the
implication is that Type Ib/c SNe come from a younger population of
progenitors. 

The connection with long-duration gamma-ray bursts has led to an increased interest in core collapse SNe
\citep{Galama98,Stanek03}.  The SNe connected with GRBs are all of Type Ic
and are massive and energetic, and tend to be more aspherical \citep{Mazzali00,Maurer10}. Observations to date suggest that only $\sim 0.1$
- $1$ per cent of SNe Ib/c produce GRBs \citep{Podsiadlowski04}, and GRB+SNe originate on average from more massive stars than SNe without GRBs \citep{Mazzali13}: typical progenitor masses are greater than $35$\Msun \ (see, e.g. \citealt{Iwamoto98,Mazzali03,Mazzali05b}). Ejecta masses have also been determined for a small number of Type Ib/c SNe that have been associated with a GRB (see e.g, \citealt{Mazzali06} and references therein). However, some
nearby GRBs that could have shown a SN apparently did not \citep{Galyam06}.
Possibly these explosions did not produce enough $^{56}$Ni to power an optically
bright SN, or there may be a difference in the stellar explosion mechanism such
that a large fraction of GRB associated SNe are underluminous. SNe Ic are also
seen in association with X-ray flashes (XRFs, \citealt{pian06}). The mass and energy of
these SNe  seem to be smaller than for GRB/SNe, pointing to a magnetar origin
of the XRF \citep{Mazzali06}.

\subsubsection{Unusual transient phenomena}

With the recent advent of the survey mode for SN searches (e.g. the Palomar
Transient Factory, Pan-STARRS, SkyMapper), we have started to uncover a wealth of transient phenomena that would
mostly be missed by the previous search mode, which was targeted mostly at bright
galaxies and had cadences optimised mostly for SNe Ia. Among the most
interesting discoveries are the long-lived superluminous SNe (SLSNe), which, as the name
suggests, outshine both SNe Ia and GRB/SNe. SLSNe come in various flavours, including events
identified as possible examples of Pair-Instability SNe (the thermonuclear
explosion of $>100$\Msun \ stars, \citealt{Galyam09}), others
that may be the product of magnetars (e.g. \citealt{Nicholl13}), or of interaction
between a SN and circumstellar material (e.g. \citealt{Smith07,Agnoletto09,Ofek10}). The true nature of these transients is still the subject of heated discussion. Another interesting discovery are the transients of various subtypes which are collectively referred to as `fast and faint'. These transients are best discovered by surveys with high cadence and good depth. An example are the Type Ib SNe away from galaxies \citep{Perets10}, which are
unlikely to be core-collapse events and are more easily linked with He explosions on the surface of white dwarfs (`.Ia SNe', \citealt{Bildsten07}) and are generally found in a luminosity-timescale space between Novae and SNe
\citep{Kasliwal10}. Several of these transients are grouped under the general name
`Ca-rich', which describes the prominence of the Ca emission lines
which develop quite early, but does not necessarily mean a high Ca abundance.
The properties of these transients are rather varied, suggesting that they
include explosions of different origin, and further data and deeper analyses are
required if we are to make sense of them. As the survey mode becomes the standard approach for transient science, more of these transients
are being observed, but events of unknown nature are also expected to be
discovered. 

\subsubsection{SNe follow-up with the LT}
\label{sec:sncurrent}

Following discovery, other facilities are employed for detailed follow-up, and the flexible scheduling capabilities of the LT makes it ideal for this task. The robotic observer enables a rapid reaction to objects which require an immediate response such as SN2011fe. For targets requiring monitoring, a monitoring timescale ranging from fractions of a second to years can be chosen in order to match the rate of evolution of the transient emission. The task of obtaining well-sampled light curves for large numbers of targets can therefore be performed with minimal interaction by the human observer. Light curves are used to distinguish between some of the sub-types of core collapse SNe; and multi-colour light curves of SNe Ia can be used to construct a bolometric light curve, which can be interpreted in terms of the decay of Ni and Co via a fairly straightforward formalism.

Spectroscopic follow-up is also a productive area of investigation. Since April 2012, the Public ESO Spectroscopic Survey of Transient Objects
(PESSTO) has shown that large scale spectroscopic classification of transients
discovered by synoptic surveys can be very productive, since this provides
large, unbiased samples of SNe. There are currently a number of programmes
active on the LT which are dedicated to the monitoring of PESSTO transients.
Large samples are necessary to track variations in the SNe population, such as
the observed differences in the light curves of SNe Ia from young, star forming
stellar populations compared to those from older, passive populations, and the
change in the ratio of these two populations with redshift.

\subsubsection{Supernova science in 2020}
\label{sec:sn2020}

Type Ia SNe are mature cosmological probes, and in the next decade they will
increasingly be used to test the temporal evolution of the dark energy equation of state \citep{Howell09}. This is in spite of the biases potentially
introduced by the unknown nature of the progenitors: for example, if there are
multiple routes to a SN Ia explosion then the mix of progenitors could
potentially change with time. The cosmological studies will usually require
telescopes of 8-metre class or larger. The role for smaller aperture facilities
is to help remove the biases through an increased understanding of the nature of
the SNe themselves, in particular elucidating the nature (or the mix) of the
progenitors. The difficulty is establishing a more statistically complete sample
of SN Ia in nearby galaxies, in order to properly characterise rare subtypes,
such as ultrabright events, intrinsically faint explosions, and SNe in unusual
environments. The same problem exists for core collapse SNe: it has been argued
that we may be missing events as close as 10Mpc \citep{Smartt09b,Thompson09}.
The core collapse population is much more heterogeneous, and so the relative
frequencies of the subtypes might be incorrectly determined with the current
sample since some subtypes are much rarer than others, which would have
implications for the parameters of the progenitor populations, such as the
metallicity dependence of rates, as well as the models of the explosions.

As discussed in Section \ref{sec:sncurrent}, there are many current projects dedicated to improving the completeness of the sample: synoptic surveys for discovery such as iPTF and Skymapper, and spectroscopic classification with spectroscopic surveys such as PESSTO. This work will continue into the next decade with the next generation of surveys, in particular LSST, which will publish of the order of $10^6$ alerts per night. The majority of these alerts will be triggered by variable stars or near Earth objects, but $1$ -- $10$ per cent being due to explosive transients is a reasonable assumption \citep{Matheson13}. Currently a large proportion of follow-up time with telescopes such as the LT is spent on photometric monitoring of SNe with the aim of providing well-sampled light curves. However the next generation of synoptic surveys will cover larger sky areas with a high cadence: LSST for example will make two visits to a given field in the same night, covering enough fields such that the entire Southern sky is imaged every $\sim$$3$ nights. The need for follow-up photometry in 2020 to complement this survey will therefore be reduced. It will not be removed entirely: the best light curve for any individual SN will be achieved with a flexible monitoring strategy which produces a higher rate of sampling during the early rise and over the peak. Another point to make with respect to LSST photometry in particular is that each visit will potentially be made using a different filter: given that the survey will be conducted with $6$ filters, the time between repeat observations in the same filter will be much longer than $3$ nights. The deep multiband imaging of facilities such as LSST will however be excellent for studies of SN host galaxies and nearby environments.

For transient science, spectroscopic follow-up will be the key role for robotic telescopes in the next decade and beyond. Systematic spectroscopic classification and follow-up will be very challenging for the transient community: even today only $\sim$$10$ per cent of PTF transients receive a classification. This is set to improve with the addition of the low resolution `SED machine' spectrograph to the instrument suite on the Palomar 60-inch telescope \citep{Ngeow13} as well as the deployment of low resolution spectrographs on other follow-up facilities, such as SPRAT on the LT \citep{Piascik14}, but the problem will be orders of magnitude greater in the LSST era. If the potential of this next generation of synoptic surveys is to be fully realised, then a huge amount of telescope time will need to be spent on spectroscopic follow-up. For classification this might see the large-scale deployment of low resolution spectrographs in order to solve the main difficulty: identification of the rarest and most interesting targets, as well as those which have been caught closest in time to the initial explosion. The primary problem is one of efficiency. In a target-rich environment one would aim to obtain a spectral classification of as many targets as possible: using a $4$-metre class telescope this would call for an $R$$\sim$$100$ spectrograph and exposure times of only a few minutes. This is comparable to the acquisition overhead of most existing telescopes and so is a very inefficient use of telescope time. The combination of a robotic observer and a fast-slewing telescope is required for efficient follow-up in the next phase of the survey era.

Following classification, scientific exploitation will require higher resolution spectrographs on telescopes of 4-metre aperture and larger. While LSST will probe deeper than the feasible limit for 4-metre class spectroscopic follow-up, there will still be thousands of potential targets per night at the bright end of the parameter space. PESSTO has demonstrated that dedicating large amounts of 4-metre time to spectroscopic follow-up can be extremely productive. Note that over the coming decades many facilities of 4-metre class and larger will be dedicating some or all of their time to highly multiplexed spectroscopic surveys using the next generation of multi-object spectrographs (e.g. WEAVE, \citealt{Dalton12}, and 4MOST, \citealt{DeJong12}). However, the design constraints of such surveys are not driven by transient science, and so the response time and monitoring cadence (if any) will not compare favourably with a single object spectrograph mounted on a telescope dedicated to targets of opportunity.

The rapid reaction capability of robotic telescopes is not just important in the LSST era for reasons of efficiency: there will also be substantial numbers of targets which demand a rapid response. The spectrum of SN2011fe was obtained by the LT $1.5$ days after the explosion, and demonstrates the benefits of this capability for SNe studies. With the next generation of detection facilities the discovery of SNe very close in time to the explosion will become much more routine, and rapid follow-up will allow study of the shock breakout and the evolution over the course of the first day. The process by which the core collapse of the star produces the shock which leads to the ejection of the stellar envelope is not fully understood, although our understanding has been advanced by the detection of several shock breakout candidates (e.g. \citealt{Campana06,Soderberg08,Gezari08b,Schawinski08,Modjaz09}) in recent years. The current difficulty is that this phase is very short, lasting $\sim$$1$ hour or less, although the duration can be extended to a day or longer if the star is surrounded by a thick wind due to mass loss prior to the explosion \citep{Ofek10,Chevalier11,Svirski12}.

\subsection{Gamma-ray bursts}

\subsubsection{Overview}

Gamma-ray bursts are the most energetic explosions to be detected in the
Universe. The first GRBs were discovered serendipitously in the late 1960s using
data from the Vela satellites \citep{Klebesadal73} and were later shown to be
distributed isotropically over the sky \citep{Briggs96}, pointing to an
extragalactic origin. GRBs can be divided into long/soft and short/hard events, depending on duration
(with the cut at $\sim 2$ sec) and spectral hardness. The former are thought to
be associated with the death of massive stars 
\citep{Woosley93,MacFadyen99}, as confirmed by
observations of nearby events (e.g. \citealt{Galama98,Stanek03,Hjorth03}) and their
interpretation (e.g. \citealt{Iwamoto98,Mazzali03,Mazzali05b}), while the
latter may be related to the merging of compact stars \citep{Eichler89,Gehrels05}.

\begin{figure}
\centering
\includegraphics[angle=0,width=1.0\columnwidth]{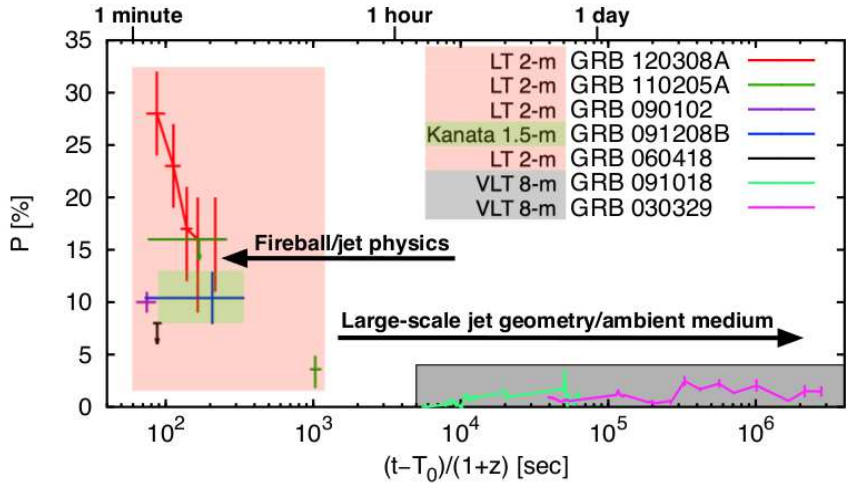}
\hfill
\caption{Rest-frame optical polarisation properties of GRBs. The degree of optical
polarisation $P$ is plotted as a function of time
after the burst in the cosmological rest frame,
$(t - T_0)/(1 + z)$, where $z$ is redshift and $(t - T_0)$ is
the time after the burst in the observer frame.
GRB 120308A, GRB 110205A, 
GRB 090102, GRB 091208B and GRB 060418 were measured at
early time. The grey shaded area shows the typical
polarisation levels of GRBs measured at late times;
representative examples GRB 091018 and
GRB 030329  are shown. Adapted from \citet{Mundell13}} \label{fig:grbpol} \end{figure}

In the 1990s the BeppoSAX observatory detected a fading X-ray afterglow to be
coincident with the long GRB 970228 \citep{Costa97}, shortly followed by a
ground based optical detection \citep{VanParadijs97}. The modern era of GRB
astronomy began in 2004 with the launch of the {\it Swift} Gamma-Ray Burst Mission
\citep{Gehrels04}. {\it Swift} continually scans a large fraction of the sky for GRB
events with the onboard Burst Alert Telescope (BAT) and rapidly and autonomously
slews after a detection to measure any afterglow with the onboard X-ray and
UV/optical telescopes. Since 2008 the Fermi Gamma Ray Space Telescope has
provided an additional source of GRB detections and has been able to measure the
gamma ray emission over a greater energy range. The afterglows of GRBs fade
extremely rapidly, and so the successes of the {\it Swift} era of GRB science have
been dependent on prompt follow-up of burst alerts by larger aperture ground-
and space-based facilities. The rapid reaction capability of the LT has
made it an extremely productive facility for GRB follow-up. The LT also benefits
from a diverse suite of instrumentation, and a real-time pipeline which makes
automated decisions as to the type of data to obtain (multicolour photometry,
spectroscopy and/or polarimetry) based on the brightness of the afterglow
\citep{Guidorzi06}. Polarimetry is an area which has been uniquely productive
for the LT, thanks to the novel RINGO series of polarimeters. RINGO has made the
first successful measurements of the optical polarisation of GRBs, which show that GRBs contain magnetized baryonic jets with large-scale uniform fields that can survive long after the initial explosion \citep{Mundell07,Steele09,Mundell13}. Figure \ref{fig:grbpol} illustrates the importance of reaction times of $\sim$$100$s or less to detect the rapidly-declining polarised emission.

\subsubsection{GRB science in 2020}

Thanks to Swift, Fermi, and comprehensive programmes of rapid follow-up, the
first decade of the 21st century has seen a tremendous advance in our
understanding of GRBs. As some of the most distant objects known, high redshift
GRBs are powerful tools with which to investigate the early Universe. However,
the number of GRBs discovered at these high redshifts is lower than was expected
prior to the launch of Swift, partly due to the fact that only $\sim$$50$ per
cent of GRBs currently receive a redshift determination. An important goal in
the coming decades will be to improve this percentage, either by increased (and
faster) spectroscopic follow up of GRB afterglows, or by calculating redshifts
from luminosity indicators in the burst light curve, which will require an
advance in our understanding of the temporal evolution in order to be
reliable. If the number of GRBs with high, well-determined redshifts could be increased,
then GRBs could potentially be used to derive reliable determinations of
cosmological parameters, as well as contribute to our understanding of
reionisation and the star formation history of the Universe \citep{Savaglio06}. One associated goal
is direct detection of the first stars in the Universe, with the identification of GRBs with Population III progenitors. The optical counterparts of high redshift bursts rapidly fade to the point where they push the capabilities of $8$-metre class telescopes to their limits. A powerful alternative is a smaller, robotic and fast-slewing telescope, which has the ability to observe the afterglow during its earlier and much brighter state.

The afterglows of GRBs of low to intermediate redshift remain accessible for longer durations, and even at this stage in the Swift era, many fundamental questions about these objects remain unanswered. About 70 per cent of GRBs
are classified as long bursts in the Swift era, with burst durations in excess of 2 seconds (see \citealt{Bromberg13} also for the short-long classification). The
observational evidence strongly suggests that long GRBs are associated with the
core collapse of massive stars (see, e.g.,
\citealt{Galama98,Stanek03,Hjorth03}).  Detailed study of the associated
SNe is critical to the advancement of our understanding of the physics \citep{Cano11},
but the number of observed associations is currently small. There
are also many open questions regarding the physical processes at work during the
initial, prompt phase of the GRB, in terms of particle acceleration and
radiation processes. High time resolution spectra of the fast spikes observed in
some GRB light curves would elucidate the internal shocks which lead to the
prompt emission. The role of magnetic fields is also an important area. Polarisation measurements have shown that early-time observations of the reverse
shock are a key diagnostic of the magnetic structure, but again, very few such
measurements have been made.

Ten years after the launch of Swift, we continue to
find unusual, and hotly debated phenomena within the GRB population. One example is the so-called
ultra-long GRBs (ulGRBs). ulGRBs are characterised by unusually long-duration prompt
emission:  GRB 111209A, for example, was active in its prompt phase for $\sim$$25,000$s \citep{Gendre13}. It has been proposed that the progenitors of events such
as this are extremely  massive stars, such as blue and yellow supergiants, to produce a sufficiently large mass-energy  reservoir to sustain the long-lived prompt emission
\citep{Gendre13,Stratta13,Levan13,Boer13}. In contrast, using a detailed statistical comparison of the high energy emission from currently known ulGRBs with the wider Swift sample, \citet{Virgili13} suggest that the current data do not yet require introduction of a distinct new class of GRB and, instead,  ulGRBs may represent the tail of the distribution of classical long GRBs, with the efficiency of kinetic-to-radiative energy conversion being a key parameter in the resultant observed prompt and afterglow properties. Extending the analysis to include X-ray light curves, \citet{Zhang14} reach a similar conclusion that the need for a new population is inconclusive.  In any case, the physical  mechanism behind the extreme energetics driving long-duration events is still poorly
understood.

Compared to the long bursts, the origin of the short bursts, which have
durations of less than 2 seconds and make up the remaining 30 per cent of the
population, is less clear. The current evidence suggests they do not have a
SN origin, but instead are associated with the mergers of two compact
objects, such as neutron stars or stellar mass black holes. This uncertainty is mainly due
to a paucity of good quality observations of short GRB afterglows, which limits
the statistical analyses. While it is clear that sGRBs are not core collapse
events, this binary merger model is not yet confirmed and the nature of the
binary components is unknown. Additionally, it appears likely that the sGRB
population is not homogeneous, with some small fraction produced by giant flares
from soft gamma repeaters: a class of neutron star which emits irregular bursts
of X- and gamma rays. The nucleosynthesis of these events, which
is predicted to produce predominantly $r$-process elements
\citep{Kasen13,Hotokezaka13} may play an important role in the chemical
evolution of galaxies. 

\subsubsection{GRB detection facilities in 2020}

Swift was launched in 2004 and Fermi in 2008. Rapid follow-up of GRB afterglows
in the next decade is predicated on the assumption that there will be a
replacement for Swift and Fermi which will be capable of detecting the
gamma ray prompt emission and providing alerts. Currently the most promising
successor is the joint French-Chinese mission SVOM \citep{Gotz09}. SVOM's
scientific payload consists of the ECLAIRs 2D coded mask imager for the
detection and localisation of gamma ray transients, and like Swift, optical and
X-ray instruments for afterglow follow-up. One of the advantages of ECLAIRs over
the BAT instrument on Swift is a greater sensitivity over a wider energy range,
down to a peak energy of 4keV, which is much softer than the 10keV limit of BAT.
Additionally, there is a possible ground component incorporated into the mission: an
array of optical cameras and two $1$-metre robotic telescopes. The purpose of
the ground component is (i) to improve on the positional information in the
burst alert, (ii) to provide a quick estimate of the photometric redshift, and
(iii) provide panchromatic optical to near-infrared light curves of the
afterglow. 

LSST is expected to begin science operations in 2021, and will have the
capability to be externally triggered by satellites such as SVOM. It is however
likely that the main LSST survey will only be interrupted in this way for the
highest redshift bursts, which are candidate probes of cosmological parameters
and dark energy, and the telescope will not react as rapidly as robotic facilities such as LT2. LSST does have the potential to provide
its own GRB triggers, with the planned observing strategy consisting of pairs of $15$ second exposures covering 
a 9.6 deg$^2$ field of view for transient detection close to real-time. Some fraction of these will potentially be orphan
afterglows: gamma ray bursts observed off axis, so the tightly collimated prompt
emission could not be detected. One potential complication will be
distinguishing GRBs from other fast-fading transients in the absence of a
detection of the high energy emission.

\subsubsection{GRB follow-up with LT2}

The rapid reaction capability of the LT is its core strength for GRB science.
Response times of $2$--$3$ min are possible because of a
robotic operator, a fast-slewing telescope, a clamshell enclosure and a rapid
response software pipeline. For a new facility the most productive improvement
would be to design the telescope such that this response time could be reduced
further. From a mechanical point of view the
challenge of meeting this requirement increases with aperture. While a larger
aperture is of course always desirable, fast-fading transients represent a case,
unusual in astronomy, in which it is not the most important parameter. Depending
on the decay rate of the transient a smaller aperture telescope can collect more
photons if it can reach the target sooner. The early-time photons may also be
more scientifically interesting: for example polarimetric observations
of the rapidly fading, reverse shock emission with the LT have opened windows
onto the nature of the relativistic fireball which are not available at later times (Figure \ref{fig:grbpol}). 

Although not currently mounted on a fully robotic telescope, the 7-channel optical/NIR GROND imager on the MPI/ESO 2.2m telescope on La Silla, Chile \citep{Greiner08} is a powerful instrument for obtaining simultaneous multicolour light curves and accurate photometric redshift determination. The ideal instrument for LT2 would have a high time resolution in order to properly characterise the highly time variable  initial stages of the afterglow \citep{Monfardini06,Greiner09,Virgili13}, and would cover optical/infrared bands from at least Sloan $r$
to $K$. The infrared bands are important for SED modelling so as to properly
characterise the extinction. Spectroscopic and polarimetric follow-up of afterglows is feasible on a 4-metre class telescope, provided the telescope is capable of extremely rapid reaction. Polarimetry of the early afterglow in particular has been a productive area for the LT \citep{Mundell07,Steele09,Mundell13}; very rapid polarisation  observations with LT2 would open the currently unexplored prompt phases.

\subsection{Multi-messenger astronomy}
\label{sec:mma}

An exciting prospect for the next decade is the beginning of the era of `multi-messenger astronomy', in which the detection of astrophysical sources will be made via non-electromagnetic means. Both neutrino and gravitational wave astronomy are expected to come to maturity and open new windows of observation on the universe. The discovery of the electromagnetic counterparts to these detections will be important for their verification and scientific exploitation, although this is complicated by a number of factors, primarily that the sky localisation of any detections is likely to be poor, and the signature of the electromagnetic counterpart is somewhat unclear.

\subsubsection{Gravitational wave astronomy}
\label{sec:gwem}

It is likely that the prospect of gravitational wave astronomy will become a reality with the advent of the advanced versions of the LIGO \citep{Abbott09,Harry10} and Virgo \citep{Degallaix13} detectors. Operations are expected to commence in 2015, with a gradual increase in sensitivity and run duration until the completion of the network in $\sim$$2022$, with the addition of the third LIGO station. The sensitivity of the Advanced LIGO and Virgo network improves on the previous versions by a factor of $\sim$$10$, with the best sensitivities for GW signal frequencies of $\sim$$100$ -- $200$ Hz \citep{Aasi13}. It is thought that the most likely detections in this frequency range will be coalescing binary systems with neutron star or black hole components. Other possibilities include `bursts' from Galactic asymmetric core-collapse SNe, and signals from the crust realignment of rapidly rotating and strongly magnetic single neutron stars (observed in the electromagnetic regime as the Soft Gamma Repeater phenomenon). The cosmic GW background is unlikely to be detected, but an upper limit could be determined in this frequency range.

\subsubsection{Electromagnetic counterparts of LIGO/Virgo sources}

The nature of the electromagnetic counterpart to the GW detection of a BNS merger is not well understood. It may be the case that most BNS mergers are not accompanied by short GRBs. Even for those that are, the prompt emission will only be detected for the small fraction in which the opening angle of the jet encompasses the line of sight. However, the majority of counterparts will be comparable to the as-yet undetected `orphan' afterglows, in which the detected emission is entirely due to the interaction of the relativistic outflow with the surrounding medium. As the observer moves further off-axis, the emission becomes much fainter and rises to peak on a longer timescale \citep{VanEerten11}. \citet{Metzger12} find that the optical emission from an on-axis event at a distance of 200Mpc decays rapidly from an initially bright $r$$\sim$$10$--$15$ (depending on the jet energy). For an off-axis event the peak of the optical emission is a day to a few days after the GW detection, and can be $r$$\sim$$18$ or fainter. They also note that if the majority of events occur in low density environments, produce low-energy jets or are not accompanied by short GRBs, then the optical afterglows will be too faint to detect. A potentially more viable counterpart in these circumstances is the redder transient powered by the radioactive decay of heavy nuclei synthesised in the neutron-rich merger ejecta: the so-called `kilonova' \citep{Metzger10}. This component rises over the course of a few days, reaching an apparent magnitude of $19.5$ to $23.5$ at peak for an event at 200 Mpc. \citet{Kasen13} used different opacities in their models and predict kilonovae are dimmer ($21 < r < 25.5$ at 200Mpc), redder and longer. Strong evidence for a kilonova was detected in the aftermath of GRB 130603B. A near-infrared component to the afterglow was detected approximately $10$ days after the Swift detection, with a colour and magnitude that are consistent with the kilonova models \citep{Berger13,Tanvir13}.

Successful detection of the counterpart will benefit from searches across the spectrum, since the interaction of the outflow with the surrounding medium will potentially produce a detectable counterpart at other wavelengths. X-ray afterglows lasting a few hours have been observed for many short GRBs. For off-axis detections \citet{VanEerten11} predict the X-ray emission will peak $\sim$$10$ days after the event. There are also predictions of radio emission on a variety of timescales, from an initial radio flare due to the excitation of the post-merger plasma \citep{VanEerten11}, through weak radio afterglows on timescales of days, to radio flares which could persist for weeks or years due to the interaction of ejecta  with the ISM \citep{Nakar11}.

As well as the gravitational wave sources  the possibility exists that there are as-yet undetected astrophysical phenomena with gravitational wave emission in excess of the detection threshold of the LIGO and Virgo detectors. The signature of any electromagnetic counterpart to such an event is of course unknown.

\subsubsection{Challenges for optical/infrared follow-up}

Observations to date with the previous generation of the LIGO and Virgo detectors have made no detections of GW emission. They have however provided upper limits on the emission from objects such as the Crab and Vela pulsars \citep{Abbott10,Abadie11}, and ruled out a merging binary neutron star progenitor for the short GRB~051103 in M81 \citep{Abadie12}. Nine event candidates were followed-up by electromagnetic facilities \citep{Abadie12b, Aasi13}. The LT was part of the follow-up programme for candidate event G23004 and demonstrated the benefits of a large aperture robotic telescope for work of this nature: the images obtained with the RATCam instrument $0.35$ days after the event provided the first real constraints on any kilonova emission from the event. Simultaneous observations were made with the wide-field `SkyCamZ' telescope which parallel points with the LT. These data were not as deep ($R$$\sim$$17$ compared to $\sim$$20$) but covered a wider field ($1^{\circ}$ vs. $4.6'$).

The uncertain signature of the electromagnetic counterpart complicates detection and follow-up, particularly when combined with the large positional uncertainty of the GW signal \citep{Nissanke13}. The median localisation of a GW detection is extremely poor ($\sim$$100$ deg$^2$) at the commencement of aLIGO operations and will still present an enormous observational challenge at full sensitivity. In 2022, it is predicted that the median localisation will be $11$ deg$^2$, with $19$ per cent of BNS mergers localised to within $5$ deg$^2$ \citep{Aasi13b}. A field of this size will contain large numbers of contaminating transients, such as near Earth objects and variable stars. These localisation estimates are made based only on the timing information, and it is possible some improvement could be made by incorporating measurements of amplitude and phase consistency between the detectors. It should also be noted that in the early years of operation the median localisation of any event varies significantly with sky position, due to the location of the detectors. Detections directly over some observing sites (La Palma, Australia) will be considerably better localised than others (Chile, South Africa). The shape of the localisation contours can also complicate follow-up: very elongated regions and detached `islands' of equal uncertainty will be commonplace. The addition of the Indian detector in 2022 will improve things, and make the median localisation relatively homogeneous with sky position.
 
Efficient coverage of the sky probability map will require the use of wide field survey facilities, such as ZTF (Zwicky Transient Facility, the successor to iPTF; \citealt{Bellm14}) or LSST, or bespoke arrays of small telescopes which can adapt the footprint of their coverage appropriately\footnote{An example is the proposed BlackGEM array of $60$cm optical telescopes, https://www.astro.ru.nl/wiki/research/blackgemarray}. The number of pointings required for conventional telescopes (with typical fields-of-view $\sim$$10'$ or less) to cover the probability map will limit their contribution to counterpart discovery, although a less time intensive approach than tiling the entire error box is to target the galaxies in the region. The success of this method will depend on the availability of galaxy catalogues which are reasonably complete up to the detection threshold. \citet{White11} provides a catalogue which is complete up to $\sim$$40$Mpc and $70$ per cent complete at $100$Mpc. The downside of this method is that it incorporates prior assumptions about the nature of the GW source. The role of conventional telescopes is more likely to be focused on scientific exploitation of the counterpart once it is discovered by wide field facilities. 

Since this is a science topic of great interest, follow-up will be very competitive and it is to be expected telescopes with aperture $8$ metres and larger will play an active role. The GRB follow-up programme with the LT has demonstrated that the rapid response of robotic telescopes enables them to explore a region of the parameter space which is beyond the capabilities of these larger aperture facilities, and given our current uncertainty as to the electromagnetic signature of any gravitational wave counterpart it may be useful here as well. Certainly the flexible scheduling capabilities of robotic telescopes make them ideal for any responsive follow-up, as evidenced by the LT's previous contributions to the gravitational wave programme. However if the counterparts are detected via a kilonova emission which rises in brightness on timescales of hours to days, the very rapid response crucial for many of our other objectives will not be as important here. Rapid follow-up is also complicated by the time for the alert to be issued by the LIGO-Virgo Collaboration. The typical latency from the event to the point where sky position information can be determined is $3$--$6$ minutes. However, at least for the initial phases of the programme there will be an additional human validation process which will add $15$--$30$ minutes. Whether this stage will be a prerequisite for alerts in $2022$ and beyond is currently undecided. The detection of any electromagnetic counterpart with the wide field facilities will also be a significantly slower process than the identification of GRB afterglows with Swift. The role for robotic telescopes may be in the counterpart discovery phase. The large positional uncertainty means each gravitational wave detection will produce a long list of candidate counterparts, most of which will be more conventional transients: supernovae, outbursting cataclysmic variables, and so on. As we discussed in Section \ref{sec:sn2020}, a fast-slewing robotic telescope with a low resolution spectrograph can address this classification problem with a high degree of efficiency.

\subsubsection{Neutrino astronomy}

Neutrino astronomy elucidates the nature and origins of high-energy charged cosmic rays. Determination of the origins of cosmic rays directly is complicated by the fact that the paths of the charged particles are perturbed by magnetic fields, in contrast to the straight paths of the neutrinos produced at the acceleration sites. Potential candidates for the particle accelerators include SN remnants, pulsars and microquasars for the lower energy rays, and active galaxies and GRBs for energies greater that $10^{19}$eV \citep{Blumer09}.

The IceCube neutrino detector at the South Pole \citep{Karle03,Icecube06} began full operations with the completion of the DeepCore array in late 2010. In 2014 construction of KM3NeT \citep{Katz06,Ulrich14} began. This is an equivalent km$^3$-scale detector for the Northern hemisphere, based in the Mediterranean sea and building on the work of the existing ANTARES detector \citep{Ageron11}. These facilities aim to detect the signatures of muons generated in charged current muon neutrino interactions. The main background for extraterrestrial neutrino events is neutrinos generated by cosmic ray interactions in the Earth's atmosphere, and so while these detectors effectively survey the entire sky, the best sensitivity is achieved for events over the opposite hemisphere where the Earth can act as a filter for atmospheric events \citep{Katz12}.

The results of an all-sky search which ran for the two years up to May 2012 were published in \citet{Icecube13}. This search found $28$ neutrino events consistent with an astrophysical origin. There is an active programme of electromagnetic follow-up of neutrino events with ground- and space-based detectors, but no counterparts have been reported to date. The challenges for electromagnetic follow-up of neutrino detections are comparable to gravitational wave follow-up, in that the expected signature of any counterpart is not clearly understood, and the uncertainty on the position is high: the angular resolution for IceCube is estimated to be around a degree \citep{Katz12}. Rapid follow-up is important, given some of the likely sources of any events. Currently triggers are issued to follow-up facilities such as the Robotic Optical Transient Search Experiment (ROTSE, \citealt{Akerlof03}). within minutes of any detection. A larger aperture robotic facility would be an important addition to this programme.

\subsection{Time domain astronomy across the electromagnetic spectrum}

In this section we give a brief overview of other facilities and surveys that will be concurrent with LT2, and will detect transient or time variable objects which will be candidates for follow-up or simultaneous observing at optical/infrared wavelengths. We also discuss in this section tidal disruption events and fast radio bursts. These phenomena have generated much interest in recent years, and for future detections rapid follow-up with robotic telescopes could be very important.

\subsubsection{The Cherenkov Telescope Array}

The Cherenkov Telescope Array (CTA: \citealt{Actis11}) will open a temporal window on the very high energy Universe. The array will consist of three types of telescope, $4$--$6$, $10$--$12$ and $24$ metres in aperture, covering the energy range from $\sim$$10$GeV to $> 10$TeV. These telescopes will be situated at both a Northern and a Southern hemisphere site, with the composition of the northern array optimised for extragalactic astronomy, and the southern array for Galactic sources. The Canary Islands are a candidate site for the northern array. The sensitivity of CTA will represent a factor $5$--$10$ improvement over contemporary Cherenkov telescopes.

The TeV sky contains a variety of different sources (see, e.g., Fig. 1 of \citealt{Hinton09}) and so the science case for CTA is diverse \citep{Actis11}, with many classes of object which would benefit from multiwavelength follow-up. GRBs are one example. CTA will only detect the brightest GRBs and so the expected rate of detection is modest ($\sim$few per year, \citealt{Inoue13}). However, extending the SED of GRBs into the CTA energy range would provide insights into the intrinsic spectrum and the particle acceleration mechanisms of GRBs which are not currently available. The CTA design plan calls for the issuing of real-time transient alerts to complementary follow-up facilities within $30$ seconds of detection, and work is underway on a real-time analysis pipeline to realise that goal \citep{Bulgarelli13}.  This capability will also be valuable for the discovery of newly flaring Active Galactic Nuclei, notably blazars. The LT currently monitors a sample of MAGIC and Fermi-detected blazars, in quiescence and in outburst, providing long-term optical light curve and polarisation monitoring for multiwavelength comparisons, the study of relativistic jet physics and particle acceleration \citep{Abramowski12,deCaneva14}.  CTA and LT2 are therefore well matched. Another example is binary stars: there are now several X-ray binaries which are known to emit radiation beyond the $100$GeV range (see \citealt{Dubus13} for a recent review). Binaries detected in gamma rays provide new opportunities for the study of particle acceleration, magnetised relativistic outflows, and accretion-ejection physics. A key probe will be the variability of the gamma ray emission with spectral state changes of the binary at lower energies, for which simultaneous multiwavelength campaigns will be required. Novae are now also recognised as potential sources of gamma rays, following the Fermi detection of V407~Cyg in 2010 \citep{Martin13} and four further novae in 2012 and 2013 \citep{Cheung13}. For V407~Cyg the gamma rays were thought to arise in shocks associated with the nova ejecta interacting with the pre-existing circumstellar wind of the evolved late-type secondary star. However, the circumstellar environments of some of the other novae are very different, and the origin of the emission is in these cases unclear.

\subsubsection{The next generation of exoplanet facilities}
\label{sec:planets}

The study of extrasolar planets (exoplanets) is one of the most rapidly advancing areas of modern astronomy. In particular, the transit method of exoplanet detection provides a great deal of scope for follow-up using instruments such as the high cadence photometer RISE \citep{Steele08} on the LT. The shape of the transiting light curve allows the determination of the orbital inclination and the relative sizes of the star and planet components. Our understanding of the structure, composition and evolution of the exoplanet population is based on the resulting mass-radius relation. A recent and productive area of investigation has been the measurement of the spectral dependence of the transit depth, which provides insights into the atmospheric composition of exoplanets  (e.g. \citealt{Charbonneau02,Knutson07,Pont08,Sing11}). At longer wavelengths detection of the secondary eclipse (e.g. \citealt{Deming05,Deming06}) provides a probe of the thermal structure of the planet, as well as constraints on the orbital eccentricity.

Exoplanet science has been considerably advanced by the Kepler mission \citep{Borucki10}, which to date has detected 961 confirmed planets and 3,601 candidates. Kepler has detected hundreds of planets of Neptune size and smaller, and has demonstrated the ubiquity of both planets and planetary systems. Analyses of the Kepler results imply that in our Galaxy there are $17$-$40$ billion Earth sized planets lying within the habitable zones of Sun-like or red dwarf stars \citep{Petigura13}. However, one of the main limitations of Kepler was that the typical visual magnitude of the host stars in the field was $13.5$ -- $14.5$. This is somewhat faint when compared to the discoveries of wider field transit surveys such as SuperWASP \citep{Pollacco06}, and so has limited the follow-up potential of ground based telescopes by comparison. The next generation of planet finding missions will seek to rectify this. For example, the Next Generation Transit Survey (NGTS), the successor to SuperWASP, will begin operations in 2014 \citep{Wheatley13}. This facility consists of twelve $20$cm telescopes based at the ESO Paranal Observatory, and will survey a sky area sixteen times the size of the Kepler field. Simulated candidate populations for NGTS and Kepler confirm the increased follow-up potential of NGTS: assuming $10$h of HARPS time per candidate, NGTS should yield $37$ Neptunes for every $7$ from Kepler. The deployment of next generation spectrographs such as ESPRESSO \citep{Pepe10} for follow-up will further increase this yield. Precise radial velocity measurements are a significant bottleneck in the confirmation and characterization of a wealth of newly discovered worlds, and $4$-metre facilities such as LT2 can contribute to these efforts by including a high-dispersion visible or near-infrared echelle spectrograph amongst its instrumental suite.

Further into the future, NASA's Transiting Exoplanet Survey Satellite (TESS: \citealt{Ricker10}) is scheduled for launch in $2017$, followed by ESA's PLAnetary Transits and Oscillations of Stars (PLATO: \citealt{Catala10, Rauer13}) mission in $2022$ -- $2024$. TESS will focus on G- and K-type stars with visual magnitudes less than $12$, and in M-type stars with visual magnitudes less than 13, and is expected to discover $1,000$ -- $10,000$ planets of Earth size or larger. The aims of the PLATO mission are similar, with a target list consisting of cool dwarfs and subgiants with $m_V < 12$, including a large sample of very bright ($m_V < 8$) stars. In addition the expected noise level of PLATO is three times lower than Kepler. The expected yield is thousands of Neptune-sized planets and hundreds of Earths and super-Earths. The number of stars surveyed, combined with the increased sensitivity, also suggests that phenomena such as exomoons, planetary rings, binary and Trojan planets could be detected for the first time. The high cadence of PLATO data will also enable detailed seismic analyses of the host stars, providing precise radii (to within $2$ per cent), masses ($4$ --$10$ per cent) and ages ($10$ per cent). The degeneracies involved in modelling transit light curves mean that this precise characterisation of the stars is essential for a detailed understanding of the properties of the planets.

Most of the atmospheric characterisation of these bright new targets will be conducted with $8$ meter class telescopes or larger, or space instrumentation such as the JWST. However there is a competitive role for $4$m class telescopes, particularly for imaging bright targets which would saturate larger aperture telescopes. Polarimetry allows for the direct detection of an exoplanet's reflected light, and is potentially a very productive area. This requires a very accurate polarimeter and a large number of photons from the target, which can be gathered in relatively short exposure times from targets in the upcoming bright planet populations provided by Kepler's K2 survey, NGTS, TESS and PLATO. Nowadays polarimetric precision down to $1 \times 10^{-5}$ can be routinely archived in $3$m class telescopes over a few hours, simultaneously in several bands (\citealt{Berdyugina08,Berdyugina11}; Piirola et al. in prep., 2014). The advantages of polarimetric programmes are that they are not time critical and can be applied to non-transiting planets. The goal is to measure variations in the polarimetric signal along the orbital phase of the planet, meaning that one can observe for a few minutes to a few hours on different dates in windows between rapid-response programmes, and interuptions due to high priority targets-of-opportunity do not result in data losses. Programmes of this nature are therefore well suited to a telescope such as LT2 in contrast to transmission spectroscopy, for which a fixed and uninterrupted observing window of several hours is required.

\subsubsection{Variable and binary stars with Gaia}

The Gaia mission \citep{Perryman01}, was launched in December 2013, and began its five year mission in July 2014, during which it will make precise astrometric measurements of one billion stars, down to a broadband ($\Delta\lambda = 440$nm) magnitude limit of $G = 20$. Additional CCDs in the image plane are dedicated to two low resolution prism spectrophotometers, enabling construction of an SED for all targets in the range $3200$ -- $10000$\AA; and high resolution ($R$$\sim$$11500$) spectroscopy over the wavelength range $8470$--$8740$\AA. The expected yield of time variable objects from this programme is extremely high: $10,000$ -- $50,000$ exoplanets, $\sim$$400,000$ eclipsing binaries, and a similar number of near Earth objects and other minor planets. In addition Gaia will image each patch of sky $\sim$$50$--$200$ times over the course of its mission, and will issue daily transient alerts, with an expected yield of $3$--$4$ SNe per day \citep{Wyrzykowski12}. LT2 will arrive too late to participate in the alerts follow-up programme (although we anticipate LT will play an active role) but the final Gaia catalogue will be published in $2020$. This catalogue will be an extremely useful resource for the elimination of contaminating variable objects during optical transient searches in the next decade, and will also contain large numbers of interesting objects for exploitation with follow-up facilities.

A systematic approach to Gaia follow-up will inevitably be limited by the sheer number of objects in the catalogue, and so the onboard instrumentation has been chosen to enable classification and parameter determination without complementary observations. However, follow-up will still be required for objects which are identified to be of particular interest, such as rare binary subclasses. Gaia parameter precision will be somewhat reduced for fainter objects: for example \citet{BailerJones10} showed that $T_{\rm eff}$ estimations made via the spectrophotometric data are reduced in precision from $0.3$ per cent at $G=15$ to $4$ per cent at $G=20$. The on-board high-resolution spectroscopy is also very limited in terms of both wavelength range and depth. The wavelength range was chosen primarily to enable radial velocity determinations using the Ca triplet. It was predicted that mission-averaged radial velocities would be determined for all objects down to $G=17$, with single-visit radial velocity measurements only available down to $G=15$ \citep{DeBruijne12}, which limits the applications to time variable objects such as spectroscopic binaries. These magnitude limits may be decreased further following the scattered light issues discovered post-launch. Similarly the cadences of Gaia photometric monitoring will be sufficient to identify eclipsing binaries and extrasolar planets \citep{Zwitter03}, but detailed modelling (e.g. \citealt{Littlefair08,Copperwheat10,Copperwheat11}) will require high-cadence follow-up observations.

\subsubsection{Tidal Disruption Events}

The phenomena of tidal disruption events (TDEs), in which a star is disrupted by tidal forces as it passes close to a supermassive black hole, has been of intense interest since the discovery of the transient source Swift J164449.3+573451 \citep{Levan11,Burrows11,Zauderer11}. The duration of the event was long compared to a GRB, with the gamma-ray and X-ray/IR emission detectable for two days and some months, respectively. The confidence in the classification of this event compared to previous tidal disruption flare candidates (e.g. \citealt{Renzini95,Bade96,Donley02}) is due to the evidence for a newly born relativistic jet \citep{Bloom11}. A second event, Swift J2058.4+0516, has since been discovered which shares many of the same properties \citep{Cenko12}. Additionally there have been recent claims of non-nuclear transients with TDE-like emission, (e.g. \citealt{Donato14}) which may be due to the disruption of a star by an intermediate mass black hole: a class of object which has thus far proved difficult to probe.

The brightest TDEs are predicted to peak at $M_V = -19$ \citep{Ulmer99}, comparable to a Type Ia SN, and so it is expected that the next generation of synoptic surveys will detect large numbers of these events. It is estimated that LSST for example will detect at least 130 events per year \citep{Lsst09,Gezari08,Gezari09}. These surveys will also catch and identify these events in their early stages, enabling rapid spectroscopic and multiwavelength follow-up to characterise, for example, the gas motions of the tidally disrupted object \citep{Lsst09}. Simulations predict that the line profiles should vary on timescales of hours to days, requiring rapid and time-resolved follow-up \citep{Bogdanovic04}. TDEs are also predicted to be strong gravitational wave sources at the moment of disruption \citep{Kobayashi04}, however, the frequency of the gravitational wave signal from such an event is likely to require a space-based detector.

\subsubsection{Radio transients}

The Square Kilometre Array (SKA: \citealt{Carilli04}) will be able to survey the sky at a rate faster than any survey telescope that has ever existed. Pathfinder facilities such as LOFAR \citep{VanHaarlem13} are already monitoring large fractions of the sky on a regular basis. One of the main science objectives of these facilities is a census of known radio transients and time variable objects such as pulsars \citep{Stappers11}, as well as the discovery of new variable radio phenomena.

A radio phenomenon of recent interest are the fast radio bursts \citep{Lorimer07,Thornton13}. These millisecond events are generally thought to have an extragalactic origin, with redshifts $z$$\sim$$0.5$--$1$, although this is still not entirely settled (\citealt{Loeb14} for example, proposed that they originate in nearby flare stars). The radio luminosity and short duration of these bursts implies a violent event, and it has been proposed that they are associated with neutron star -- neutron star mergers \citep{Pshirkov10}, in which case these bursts might be an electromagnetic counterpart to gravitational wave events detectable by LIGO. Other possible causes are magnetic reconnection in a neutron star magnetosphere \citep{Somov11} or SNe explosions in a binary system impacting a neutron star magnetosphere, \citet{Egorov09}, and recently \citet{Falcke14} proposed that these bursts might be evidence of a new formation channel for stellar mass black holes that are not seen as GRBs, via the collapse of massive rotating neutron stars. To date there has been no detection of a counterpart to these bursts at wavelengths shortwards of the radio. Optical follow-up will be important for the elucidation of these sources: if they are caused by a neutron star -- neutron star merger then they should be detectable as an `orphan' GRB. For their neutron star collapse model \citet{Falcke14} propose that optical and X-ray data could constrain the emission process, baryon load and delay of the collapse. The flexible scheduling capabilities of robotic telescopes provides scope for frequent and efficient programmes of simultaneous optical and radio observations.

\subsubsection{Minor Planets and Comets}

A key goal of current and future synoptic surveys is to provide a census of solar system objects. The detection of potentially hazardous objects is the primary purpose of Pan-STARRS for example, and is a key driver of the telescope design and observing cadences of the LSST project \citep{Ivezic07}. ESA also has a mandate to pursue a Space Situational Awareness Programme in order to safeguard European commercial activities in Earth orbit\footnote{http://www.esa.int/Our\_Activities/Operations/\\Space\_Situational\_Awareness}. The LT has been a popular tool for observation of minor planets and comets, with the database of the IAU's Minor Planet Centre\footnote{http://www.minorplanetcenter.net} recording over 500 LT observations since January 2009. One science highlight is the first detection of the Yarkovsky-O'Keefe-Radzievskii-Paddack (YORP) effect: a torque caused by solar radiation pressure and the recoil effect from anistropic emission of thermal photons \citep{Lowry07}, which can modify the rotation rate of minor planets. As our synoptic survey capability improves, interest is increasing in small ($< 100$m) objects. Objects of this size still present a significant hazard: it is estimated that the asteroid which caused the 1908 Tunguska event was $\sim$$30$m in size for example \citep{Chyba93}, but it is believed only $1$ -- $2$ per cent of this population has been observed \citep{Brown13}. These objects are only detected in surveys while they are close, and usually don't spend much time ($\sim$days) in near-Earth space. They therefore require timely follow-up to extend the observation arc before they are too far away to observe: perturbations mean many observations are needed within about the first week otherwise the asteroid is not found again. The LT is capable of tracking targets moving at non-sidereal rates of more than 2 arcminutes per minute, and we anticipate LT2 will have a similar capability. Fast non-sidereal tracking allows spectroscopic observations of solar system objects, which is an important tool for taxonomy (see, e.g., \citealt{Bus02,Lazzarin04,Urakawa13}). Interest in the physical and chemical properties of comets in particular is likely to be enhanced by ESA's ROSETTA mission \citep{Glassmeier07}, which began its close-up study of comet 67P/Churyumov-Gerasimenko in 2014.

\subsection{The unknown}

\begin{figure}
\centering
\includegraphics[angle=270,width=1.0\columnwidth]{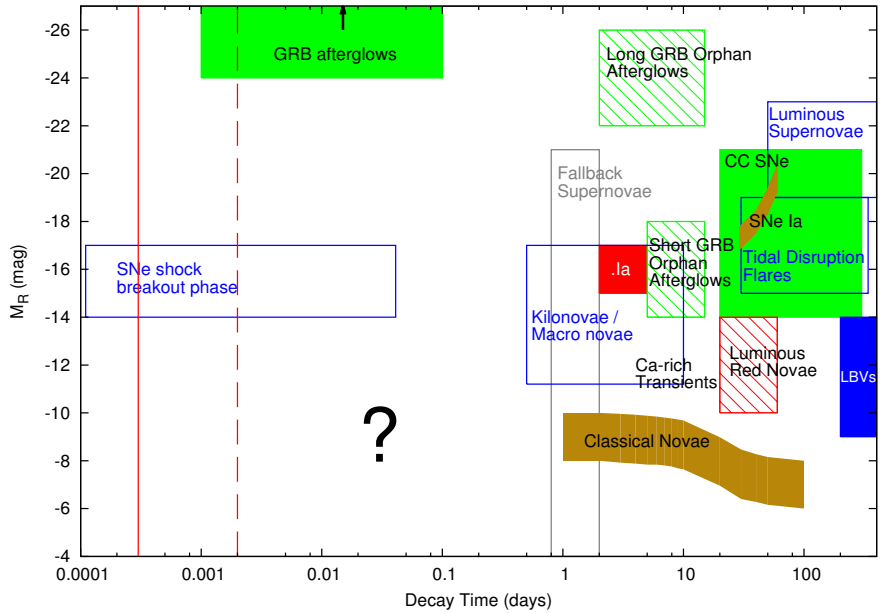}
\hfill
\caption{Explosive transient phase space. This is a modified version of a figure which appears in the LSST Science Book \citep{Lsst09}, which in turn was a version of the original figure in \citealt{Rau09} modified to show the unexplored sub-day phase space which will be probed by the next generation of synoptic surveys. We include here the SNe shock breakout phase as the example of a phenomenon which will continue to be of great interest and will require a sub-hour follow-up response. The vertical red lines denote for reference the typical response time of the LT (dashed) and LT2 (solid).} \label{fig:transients} \end{figure}

The next decade will see a vast increase in the capabilities of survey facilities, in terms of sky coverage, cadence and wavelength, revealing new transient and variable objects at an unprecedented rate. As well as the transient classes we already know of and have discussed in the previous sections, the expectation is that these surveys will uncover new and exotic time variable phenomena. This is due to the opening of the time domain window on new electromagnetic and non-electromagnetic regimes, and also the exploration of the faint/fast region of the transient magnitude / timescale phase space (Figure \ref{fig:transients}, adapted from the version of the \citealt{Rau09} figure which appeared in the LSST Science Book). The transient phenomena in these unexplored regions of parameter space are potentially as plentiful and diverse as the array of subtypes displayed on the right-hand side of this figure, and it is generally the case that when a new window is opened on the Universe, it is the unanticipated phenomena which provide the greatest scientific impact. As with the transients of known type, there will be a pressing need for follow-up facilities, in particular follow-up spectroscopy, to classify and exploit these new discoveries. LT2, planned to be the largest aperture robotic telescope in operation, provides a uniquely powerful capability with which to pursue targets in the faint/fast regime. We note that LT2 is well placed to coordinate with LSST for rapid response follow-up, since both LT2 partner institutions intend to join the LSST project (LJMU as part of the UK membership, and the IAC as an individual institute).

\subsection{Summary of science requirements}
\label{sec:scisummary}
The study of the time domain has fundamental importance in a wide range of astrophysical contexts, and we are on the brink of a new era of discovery with many major new facilities close to deployment. In this section we have sought to explore the role for a new robotic telescope by examining the science topics and facilities which would benefit from complementary robotic observations. The LT has been a very successful facility for time domain astronomy and we intend LT2 to build on this success by identifying and improving the capabilities which have made the LT such an effective tool.

\subsubsection{Main scientific objectives}
\label{sec:objectives}

\begin{enumerate}
\item The key advantage of robotic observing is the rapid response capability for targets of opportunity. It is therefore no surprise that many of the LT's most important scientific findings have been in the field of transient science: Novae, SNe and GRBs. The first ten years of LT operations coincided with the birth of the survey mode for transient astronomy. The yield of the next generation of survey facilities such as LSST promises to be enormous, and the most pressing need will be follow-up spectroscopy for classification and exploitation. Maximally exploiting this target-rich environment in order to uncover the rare subclasses of objects requires extensive and efficient follow-up by dedicated telescopes.

\item GRB science has been one of the cornerstones of the last decade of time domain astronomy, thanks to the influence of the Swift and Fermi satellites. The fast-fading nature of the afterglows means robotic telescopes are ideal tools for follow-up observations, since a telescope with a rapid reaction capability can collect more photons than a slower-slewing telescope of much larger aperture (e.g. \citealt{Mundell13}). The LT has been very successful in this field: to build on this success with LT2 requires not just an increased aperture, but also a more rapid response in order to probe even closer in time to the initial explosion. GRB science in the next decade is predicated on the existence of a Swift successor in orbit to provide alerts, but it is not the only science case requiring a rapid response. Many new survey facilities will provide near real-time alerts with rapid follow-up in mind. Facilities such as CTA and SKA will also open the temporal window on previously unexplored wavelengths. 

\item The detection and follow-up of counterparts to gravitational wave sources is a major goal for the time domain community in the coming decade. The nature of the counterpart is unclear, and while the rise time of any kilonova emission will not warrant the very rapid response demanded by some of our other objectives, the advantages of robotic telescopes for gravitational wave follow-up have been demonstrated by the historic contributions of the LT to the LIGO programme. Given the huge positional uncertainty in any detection the list of candidate counterparts will be large, and the problem of efficient classification of those candidates is similar to the challenge of maximising LSST exploitation.

\item As well as rapid reaction, a key strength of robotic telescopes is their flexible scheduling capabilities. It is a simple task for a remote observer to match the cadence of monitoring observations to the variability timescale of the object of interest, be it fractions of a second or years. This makes robotic telescopes powerful tools for the study of periodically variable objects, such as binary stars and transiting exoplanets. Facilities such as Gaia and the next generation of planet finders will provide huge numbers of potential targets for follow-up. The remote scheduling capabilities of robotic telescopes means large follow-up programmes of such objects can easily be combined with a programme of rapid response to transient alerts. Of particular interest are programmes consisting of relatively short ($\sim$$1$h) observations, such as polarimetric measurements of exoplanets (as described in Section \ref{sec:planets}), since they can be easily interrupted by targets-of-opportunity. 
\end{enumerate}

\subsubsection{Key technical drivers}
\label{sec:drivers}

The objectives enumerated in Section \ref{sec:objectives} provide three key drivers for the telescope design, which we list here.

\begin{table}
\caption{Summary of some of the different transient phenomena we intend to pursue with LT2, and the typical response time we would aim to achieve}
\label{tab:response}
\begin{center}
\begin{tabular}{lll}
Response time & Transient phenomena\\
\hline
$< 1$ min after trigger & GRB afterglows (inc. orphan afterglows\\
			  & and on-axis GW counterpart) \\
			  & Fast radio phenomena \\
			  & SNe shock breakout \\
			  & Neutrino events \\
			  & High energy (CTA) transients \\
$< 1$ hr after trigger & Early time SNe \\
$\sim$hours  	  & Novae\\
		  & SNe detected closer to peak\\
$\sim$day -- days & GW kilonova emission\\
		  & Tidal disruption events\\
\hline
\end{tabular}
\end{center}
\end{table}

\begin{description}
\item[{\bf Rapid reaction}:] LT2 will be a fully robotic telescope, as the LT has demonstrated the power of robotic operations for the study of temporal variability and targets of opportunity. The LT has shown that the rapid reaction provided by robotic operations can be crucial for the study of fast-fading transients such as GRB afterglows. Indeed, for advancing our GRB follow-up programme reaction time is more important than anything else: even an increased aperture. The next decade will see more classes of transients which demand a rapid response: we list some of these in Table \ref{tab:response}, along with the typical response times we would aim for with LT2. In some cases our response will be limited by the capabilities of the triggering facilities. In other cases (such as for gravitational wave counterparts) the optimum response time is unclear due to the uncertain signature of the electromagnetic counterpart. Rapid response also maximises our ability to catch the unknown, which is potentially of great importance as we probe the time variable sky at shorter cadences and new wavelengths. We note again that the LT was not originally designed with GRB science in mind, but with the launch of Swift its response capability proved most serendipitous.

There is a second aspect to the case for rapid reaction, which is arguably even more important. The target-rich environment of the LSST era represents a fundamental change to the field of transient astronomy. Even today the rate of detections is far too great for spectroscopic classification programmes to keep up: in the next decade this problem will be orders of magnitude greater. Without large scale programmes of follow-up the majority of the most rare and unusual objects observed by LSST will not be picked out and much of the potential of the survey for the time domain community will be squandered. Automated brokers will provide some help with preliminary classifications but the list of targets worthy of a spectrum will still be extremely large. Existing spectrographs such as SPRAT have shown that short exposures and low resolutions are all that is required for this work, but with the typical telescope overhead for changing target means a programme like this will be extremely inefficient on existing telescopes. A large scale programme of transient follow-up therefore demands an extremely fast slewing telescope to maximise the science gain, even when the targets themselves are varying on longer timescales.

\item[{\bf Aperture}:] The transients discovered by the next generation of synoptic surveys and the optical/infrared counterparts will be significantly fainter than those typically observed by the LT today, and since the primary observing mode of the new telescope will be spectroscopic rather than photometric, a larger aperture than LT will be required. The LSST camera for example saturates at a magnitude of $\sim$$16$ and so there will be very limited opportunities for spectroscopic follow-up with $2$-metre class telescopes. The aperture requirement has to be balanced with our requirement for rapid reaction since what can be achieved in terms of slew speed is dependent on the telescope mass, which is heavily influenced by the size of the primary mirror.

\item[{\bf Instrumentation}:] The cadences of future synoptic surveys means that the role for photometric follow-up will be somewhat diminished, but follow-up telescopes with a spectroscopic capability will be very much in demand. A low resolution ($R$$\sim$$100$) for rapid classification will be required, as well as higher resolutions ($R$$\sim$$1000$ or greater) for exploitation of the targets which need it. However, much of the success of the LT has been due to its diverse and flexible instrument payload, and a number of our science goals require polarimetry or high cadence imaging. We therefore consider the capability to simultaneously mount and automatically change between a number of different instruments a key driver for the project.

\end{description}

\section{Telescope design}
\label{sec:telescope}

In Section \ref{sec:drivers} we identified the three main technical drivers of the project: a very rapid response capability, a increased aperture compared to LT and a diverse instrument suite with a focus on spectroscopy. In this section we discuss these three drivers in more detail and present some results from the preliminary design studies which we are currently undertaking.

\subsection{Rapid response capability}

The science case for LT2 calls for a world-leading response capability. In this section we explore this requirement. As well as discussing the telescope slew time, we also consider pointing accuracy and the enclosure, both of which can be the limiting factor in target acquisition time.

\subsubsection{Telescope slew time}
\label{sec:slew}

\begin{figure*}

\centering
\includegraphics[angle=270,width=1.0\textwidth]{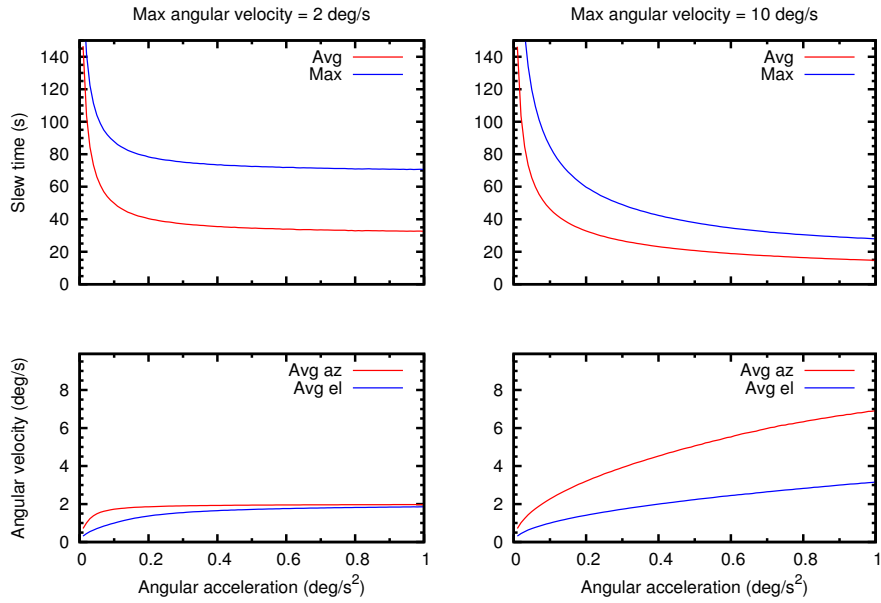}
\hfill
\caption{Simple slew models for an alt-az telescope. These plots are the results of a Monte Carlo simulation in which a telescope with a fixed maximum angular velocity slews from a random position on the sky to a second random position. The telescope accelerates and decelerates at the same, constant angular acceleration. {\bf Top}: Slew time is plotted against angular acceleration. The red line denotes the average slew time and the blue line denotes the longest possible slew time. {\bf Bottom}: Angular acceleration plotted against the average {\it maximum} angular velocity of the slew. The red line denotes the velocity of the azimuthal axis and the blue line the velocity of the altitude axis. {\bf Both}: The maximum angular velocity is fixed at $2$deg/s in the left panels and and $10$deg/s in the right panels. } \label{fig:slew} \end{figure*}

Our target for response time is that the telescope would be on average, able to start obtaining data within $30$ seconds of receipt of an electronically transmitted alert from another facility. This time incorporates the telescope blind pointing time, the mirror settling time and any mechanical movement of the enclosure (Section \ref{sec:enclosure}). This response time is significantly in advance of what is achievable with current facilities. We have made some simple calculations to investigate the feasibility of this goal, and in Figure \ref{fig:slew} we plot the results of some Monte Carlo simulations which assume a simple model of a telescope slewing to random points in the visible sky, in which the telescope accelerates at a constant rate to a maximum angular velocity, and then decelerates at the same rate. The plots show the results for different values of angular acceleration for two cases of maximum angular velocity: $2$ deg/s and $10$ deg/s. For each velocity there are two plots: one showing the average and maximum slew time, and one showing the average top speed of the slews in the azimuth and elevation axes. 

The case where the maximum angular velocity is $2$ deg/s approximates the existing LT. The angular acceleration of the LT is approximately $0.2$ deg/s$^2$ and we see that most slews reach the maximum velocity: in other words, the slew time of the LT is limited by the maximum angular velocity parameter, and increasing the acceleration would provide little improvement. However, in the case when the maximum angular velocity is large, the reverse is true: slews are never long enough to reach this maximum velocity, and so slew time is dominated by the acceleration parameter. For this simple model the maximum velocity can be relatively modest at around $4$ deg/s as long as the acceleration is at least $0.4$ deg/s$^2$. Our estimation is that this is physically achievable, although the ease at which it will be achieved is dependent on the moment of inertia of the telescope. The fast slewing requirement therefore is the main driver for the telescope design, influencing the choice of primary and secondary mirror construction and the focal ratio (Section \ref{sec:optics}), the choice of materials for the telescope structure, and the location of the focal stations (Section \ref{sec:instrumentation}).

\subsubsection{Pointing and tracking}
\label{sec:point}

The response capability is also dependent on the accuracy and precision of telescope pointing. Our pointing accuracy is driven by the need to robotically acquire a target for spectroscopic follow-up. It is currently undecided as to whether the LT2 spectrograph will use a long slit or integral field unit, but here we assume a slit as this imposes the strongest constraint. We derive a baseline requirement for pointing precision in operation of $0.3''$ RMS by assuming a long slit spectrograph with a $1.0''$ slit, a point source with FWHM $0.8''$, and then calculating the maximum offset between the slit centre and the point source at which $70$ per cent of the light still passes through the slit. This level of precision will be extremely difficult to achieve with the initial blind pointing, and so we choose to impose this precision requirement for a telescope movement of the size of the field of view, providing the initial slew places the object within the field of view. With respect to our target of a $30$ second response time, our aim therefore is to put the target within the field of view for photometric measurement within this time. The time to position the target for spectroscopic observation will be additional to this, and will largely depend on the integration time for the acquisition image. Of course any acquisition images would potentially be useful science data themselves.

One of the larger components of the overhead budget for current observations is guide star acquisition. We therefore consider excellent open loop tracking to be an important component of the design, to give observers the flexibility to obtain unguided exposures when the most rapid response is required. Our design requirement is that the tracking must allow for a monochromatic exposure of at least ten minutes with an image elongation of no greater than $0.2''$. The closed loop tracking must allow for a guided monochromatic exposure of one hour with an image elongation of no greater than $0.2''$.

\subsubsection{Enclosure}
\label{sec:enclosure}

The acquisition time of contemporary telescopes is often limited by the rate of rotation of the dome, rather than the slew time of the telescope itself. The novel clamshell design of the LT enclosure negates this problem by giving the telescope an unencumbered view of the entire sky, and has the additional benefit of eliminating the phenomenon of dome seeing. We would anticipate the enclosure for LT2 will be of a similar design.

\subsection{Optics}
\label{sec:optics}

\begin{figure*}
\centering
\includegraphics[angle=270,width=1.0\textwidth]{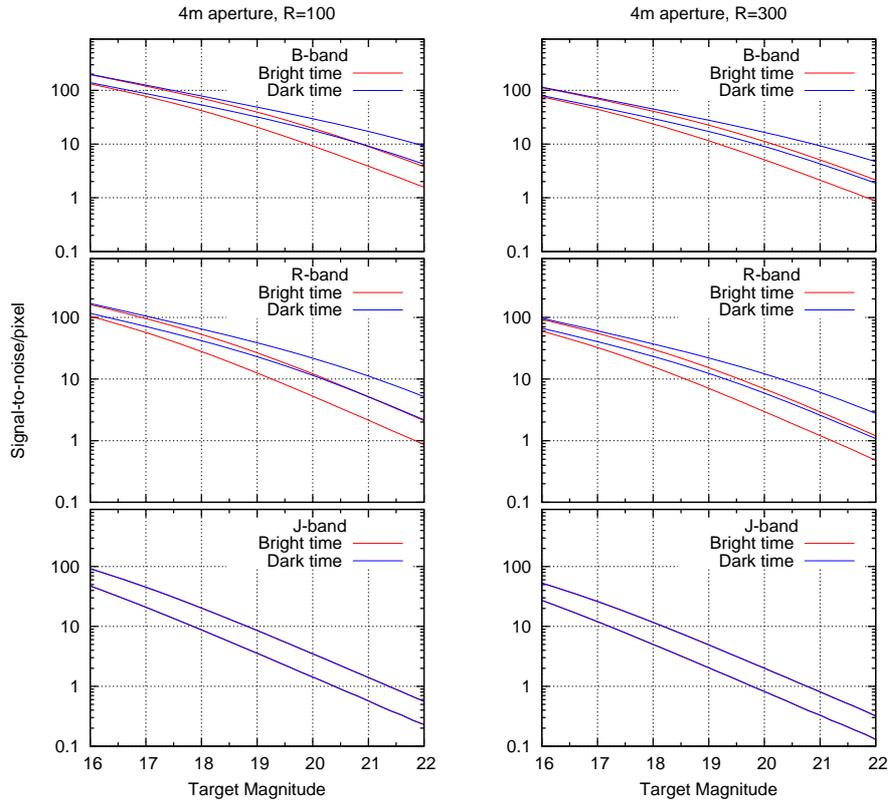}
\hfill
\caption{S/N estimates for an $180$s integration using a typical long slit spectrograph mounted on a $4$-metre aperture telescope, in the $B$- $R$- and $J$-bands. We assume typical sky brightnesses for La Palma (by using values in the ING SIGNAL calculator; http://catserver.ing.iac.es/signal/), a throughput (comprising atmosphere, telescope, instrument and grating) of $20$ per cent, a chip quantum efficiency of $80$ per cent, a read noise of $4$ electrons, a plate scale of $0.4''$/px and a slit width of $1.0''$. The blue and red lines show the S/N in bright and dark time respectively, for object point spread functions of $0.5''$ (upper) and $1.5''$ (lower). The left and right panels are for grating resolving powers of $R=100$ and $R=300$, being representative of the typical resolutions which would be employed for transient classification.} \label{fig:sn1} \end{figure*}

\begin{figure*}
\centering
\includegraphics[angle=270,width=1.0\textwidth]{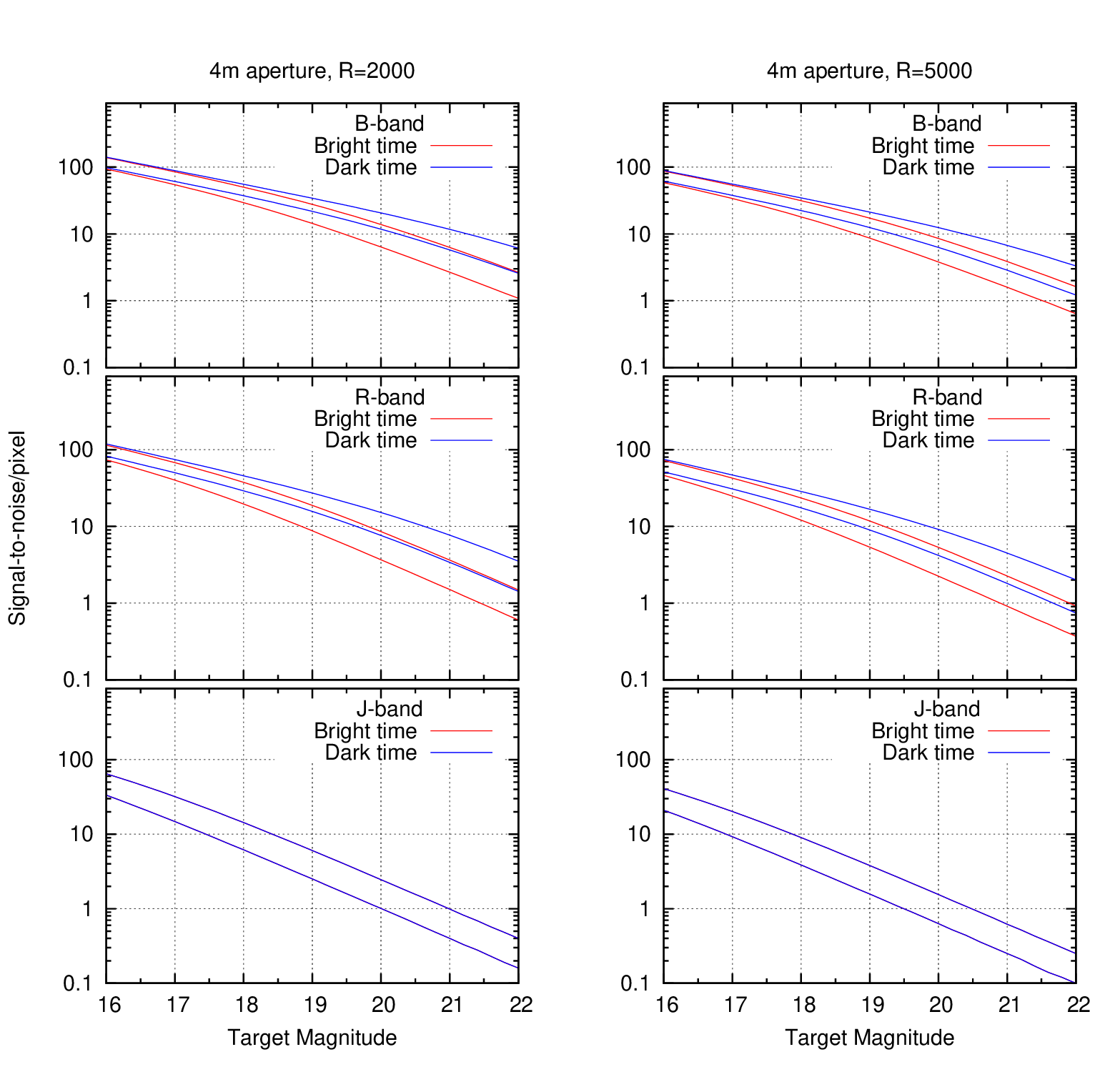}
\hfill 
\caption{S/N estimates using a typical long slit spectrograph mounted on a $4$-metre aperture telescope, as in Figure \ref{fig:sn1}, but for a longer $1800$s integration with grating resolving powers of $R=2000$ (left) and $R=5000$ (right). This exposure time and resolution is more typical of what might be used for a science spectrum of a source of known type.} \label{fig:sn2} \end{figure*}

We wrote a simple signal-to-noise calculator for a long slit spectrograph, and made some reasonable assumptions for instrumental throughput and observing conditions. We then estimated the signal-to-noise in various bands for different combinations of telescope aperture and spectrograph resolving power, using some reasonable assumptions. Based on our science case we anticipate our typical spectroscopic targets will have visual magnitudes of $18$ or more:
LSST for example will report transient detections over the magnitude range $r = 16$ -- $24.5$. We find an effective aperture of at least $4$-metres is required for efficient follow-up. In Figures \ref{fig:sn1} and \ref{fig:sn2} we plot predictions of signal-to-noise for this aperture. Figure \ref{fig:sn1} assumes a short exposure time of $180$s and grating resolving powers of $R=100$ and $R=300$. This resolution is sufficient for a transient classification spectrograph: for example the SED machine \citep{Ngeow13} and SPRAT \citep{Piascik14} have resolving powers $R$$\sim$$100$ and $R$$\sim$$350$, respectively. Figure \ref{fig:sn1} assumes an exposure time of $1800$s and grating resolving powers of $R=2000$ and $R=5000$, which is a resolution which is generally sufficient for the scientific exploitation of our targets of interest.

As we noted in Section \ref{sec:slew}, the optical design of the telescope is driven by our fast-slewing requirement. Our preliminary studies recommend a Ritchey Chr\'etien design, with an f/1.0 or f/1.5 primary mirror. The recommended final focal ratios would be f/6.5 -- f/8 for the f/1.5 primary, or f/7 -- f/10 for the f/1.0 primary. These two alternatives take a different approach to the problem of minimising the moment of inertia of the telescope: in the first case the overall length of the system is reduced, and in the second the longer final focal length allows for a comparatively smaller and less massive secondary mirror.

\begin{figure}
\centering
\includegraphics[angle=0,width=1.0\columnwidth]{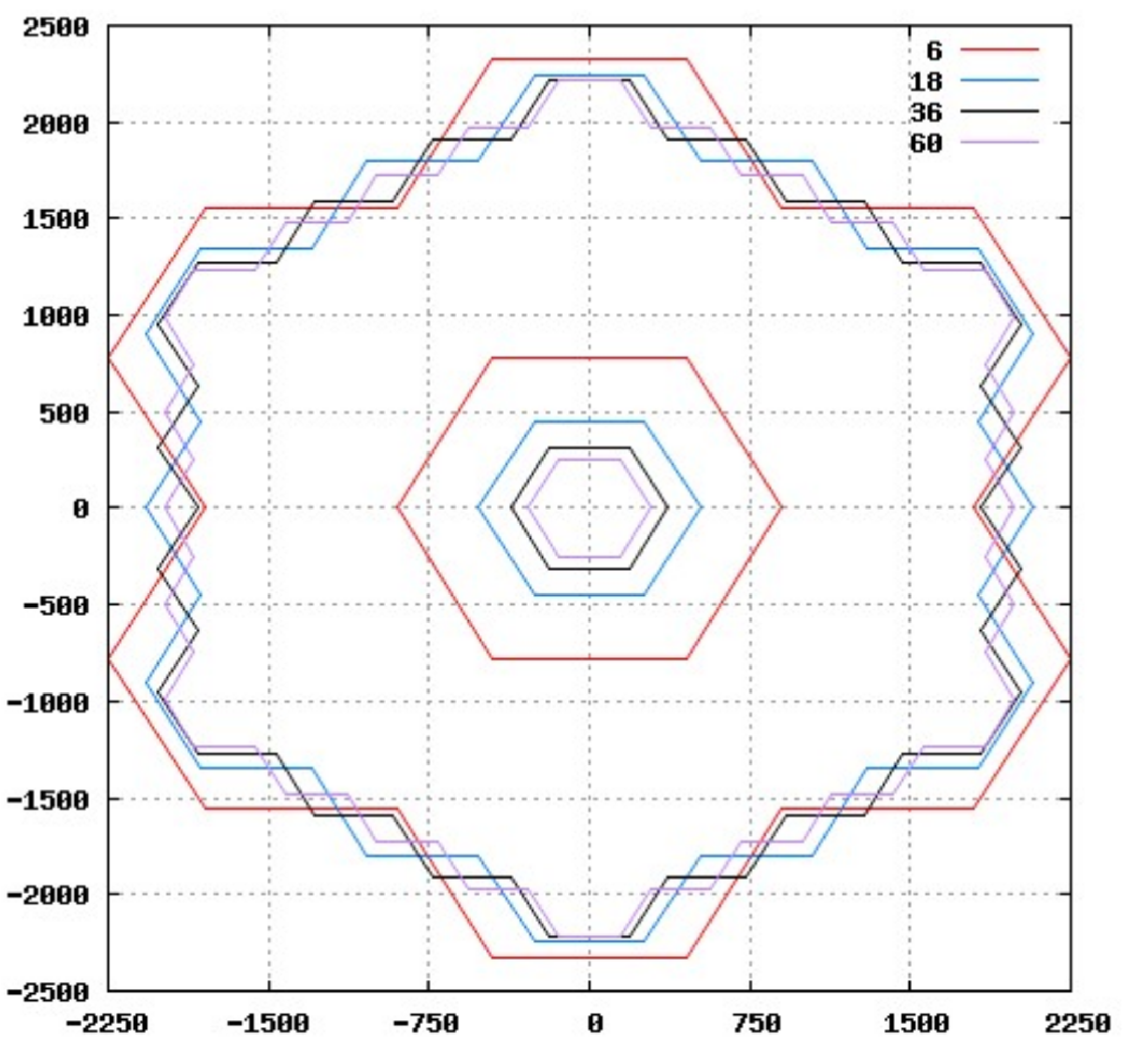}
\hfill 
\caption{Projected mirror apertures for segmented primary mirrors with 6, 18, 36 and 60 hexagonal tiles. All four apertures have the same area: $4\pi$m$^2$.} \label{fig:segments} \end{figure}

Reducing the mass of the primary mirror is potentially key to meeting our fast-slewing requirement. A monolithic meniscus with a $4$-metre diameter would weigh $\sim$$5500$kg. We are therefore considering segmented solutions for LT2, and it seems highly likely at this stage that this will be preferred. In Figure \ref{fig:segments} we plot the projected mirror apertures for primary mirrors consisting of different numbers of hexagonal segments. As the number of segments is increased the overall weight of the mirror goes down. A $6$ segment mirror would have a total mass of $\sim$$2000$kg for example, which is reduced to $\sim$$850$kg for a $36$ segment mirror: this is lighter than the LT's $2$-metre diameter monolithic mirror. The reduction in mirror mass is compensated to some extent by the mirror support structure and systems, which increase in complexity and mass with the number of segments. The problem of segment alignment also becomes much more difficult with more segments. Of fundamental importance is the impact the different segmented designs will have on image quality, which we are currently investigating. We discuss the various segmented options for the LT2 mirror in more detail in Section 4 of \citet{Copperwheat14}.

\subsection{Instrumentation}
\label{sec:instrumentation}

As discussed in Section \ref{sec:science}, the most pressing need in the next decade will be for spectroscopic classification and follow-up. We therefore intend the main LT2 instrument to be an optical/infrared spectrograph capable of low to intermediate dispersions (up to $R$$\sim$few thousand). The wavelength range of the spectrograph is to be decided, although we would aim to push as far into the infrared as possible to facilitate the study of extragalactic transients. Our design specification for the telescope itself imposes an effective wavelength range of at least $350$nm to $2.0\mu$m, covering the optical B-band to the infrared H-band. The choice of a slit or integral field unit (IFU) for the spectrograph will be determined by throughput: a high throughput is generally more straightforward to achieve for a long-slit spectrograph, although a slit imposes a tougher constraint on pointing precision than an IFU (Section \ref{sec:point}), since the reliability of automatic acquisition routines is of course vital for robotic operation. However, science operations with SPRAT \citep{Piascik14} on the LT in late 2014 have demonstrated that reliable acquisition is possible with a robotic long-slit spectrograph. 

As well as spectroscopy, a number of our intended goals would be facilitated by other instrumental capabilities, such as high-cadence imaging or polarimetry. The diverse instrument suite of the LT is one of its core strengths: all instruments are mounted simultaneously, and instrument changes can be made in the middle of a night with a typical overhead of $30$s. LT instrumentation can therefore also be fairly specialised and relatively simple, which means the lead time from instrument concept to science operations is quite low. This has enabled the LT to respond quickly to new and evolving scientific needs. We would like LT2 to have these same strengths, and therefore a requirement of the telescope design is the ability to mount up to five instruments simultaneously, with the capability for automatic changes.

\section{Site}
\label{sec:site}
Our preference is to co-locate LT2 with the LT at the Observatorio del Roque de Los Muchachos on the Canary island of La Palma, Spain; and LJMU is developing this option in partnership with the Instituto de Astrof\'isica de Canarias (IAC). The ORM is one of the best observing sites in the world, with low levels of light pollution, median seeing of $0.76''$\footnote{http://www.iac.es/proyecto/site-testing/index.php}, and more than 80 per cent of nights are photometric. There are obvious logistical advantages to choosing the ORM for LT2: it is a familiar environment and a relatively short trip for Liverpool-based staff in the event of an equipment malfunction. However, the most important consideration for the site choice is how well it matches our science requirements. In particular, given that LT2 is designed for the scientific exploitation of new transients discovered by survey facilities, the distribution on the sky of those transient detections is crucial. Space-based facilities such as SVOM will detect potential targets over the entire sky, and so do not constrain site choice. This is also true of gravitational wave triggers from aLIGO, although the sensitivity and localisation will vary somewhat with sky position. At extreme high energies, CTA will have both northern and southern sites, with the composition of the northern array optimised for extragalactic work, and the southern array optimised for the study of Galactic sources. Given that one of our primary synergies with CTA is the study of GRBs, we would prefer a northern site for CTA follow-up with LT2.

It is anticipated that the majority of LT2 time will be spent following up SNe discovered with all-sky surveys. This science requirement does not strongly favour one hemisphere over the other. From the Canaries a telescope can cover the whole Northern hemisphere and an important part of the Southern hemisphere, which means that collaborations with most of the new state-of-the-art Southern facilities are feasible. Currently there are large survey facilities operational in both hemispheres (e.g. iPTF, Skymapper), and SNe are of sufficient interest that this will certainly still be true in the next decade. Indeed, the era of gravitational wave astronomy is likely to bring new synoptic facilities into operation since counterpart detection will require high cadence wide field imaging: examples of proposed facilities include BlackGEM in the south and GOTO in the north. However, it is undeniable that the most important survey facility of the next decade will be LSST, situated at Cerro Pach\`on, Chile.  The majority of LSST time will be expended on the primary `wide-field-deep' survey, which will image the entire visible sky with an airmass less than $1.4$ every three nights on average. This airmass limit, chosen to fulfil the image quality requirement, means that the declination range of the survey will be $-75$ to $+15$ degrees. The latitude of La Palma is $+28.8$ degrees, so there is still significant overlap with the LSST field, even from a northern site. For the purposes of illustration, a target with a declination of $-30$ degrees is visible from La Palma for over $1.5$ hours at an airmass of $2$ or better. A target with a declination of $-10$ degrees is visible from La Palma for over $6.5$ hours at an airmass of $2$ or better, and $4$ hours at an airmass of $1.5$ or better. There is therefore considerable scope for follow-up of LSST transients from La Palma. This is reinforced by the sheer number of alerts it is expected the LSST project will issue. The majority of these will be near-Earth objects and variable stars, and many of the remaining transients will be too faint for 4-metre class spectroscopic follow-up, but a pessimistic estimate of the number of potential targets is still of the order of $\sim$$100$s per night. The challenge, even for northern facilities, will be selecting the targets with the most scientific potential from a very long list of candidates.

\section{Future of the LT}
\label{sec:ltfuture}

Locating LT2 on La Palma offers a couple of possibilities for the future of the LT. We would ideally like to keep the LT in use, but since the majority of staff effort will shift to the new telescope, we would likely aim to simplify LT operations by running it as a single instrument facility. Since we anticipate LT2 will focus on spectroscopic observations, an obvious role for the LT is to provide complementary and simultaneous photometry. As part of the LT2 project we are considering developing a new, prime focus imaging camera for LT, the field of view of which would be approximately $2 \times 2^{\circ}$. The LT could then serve as our own survey facility, and we would run the two telescopes together along the lines of the PTF model. Run as a discovery facility the LT could detect much fainter transients than the $1.2$ metre Schmidt telescope at Palomar, due to both the increased aperture and the significantly better median seeing at the ORM, at the price of a much smaller field of view. By controlling our own survey telescope we reduce the time delay inherent in reliance on external triggers (which varies in importance dependent on the triggering facility). A $2$ metre telescope with this field of view would also be able to make a contribution to the discovery of counterparts to poorly localised triggers, such as gravitational wave events (Section \ref{sec:gwem}). Keeping the LT operational as a dedicated imaging facility would mean that it could continue to be used for the educational programme (Section \ref{sec:edu}), although the design of the new camera would need to account for the fact that the targets in this programme tend to be significantly brighter than typical science targets.

\begin{figure}
\centering
\includegraphics[angle=0,width=1.0\columnwidth]{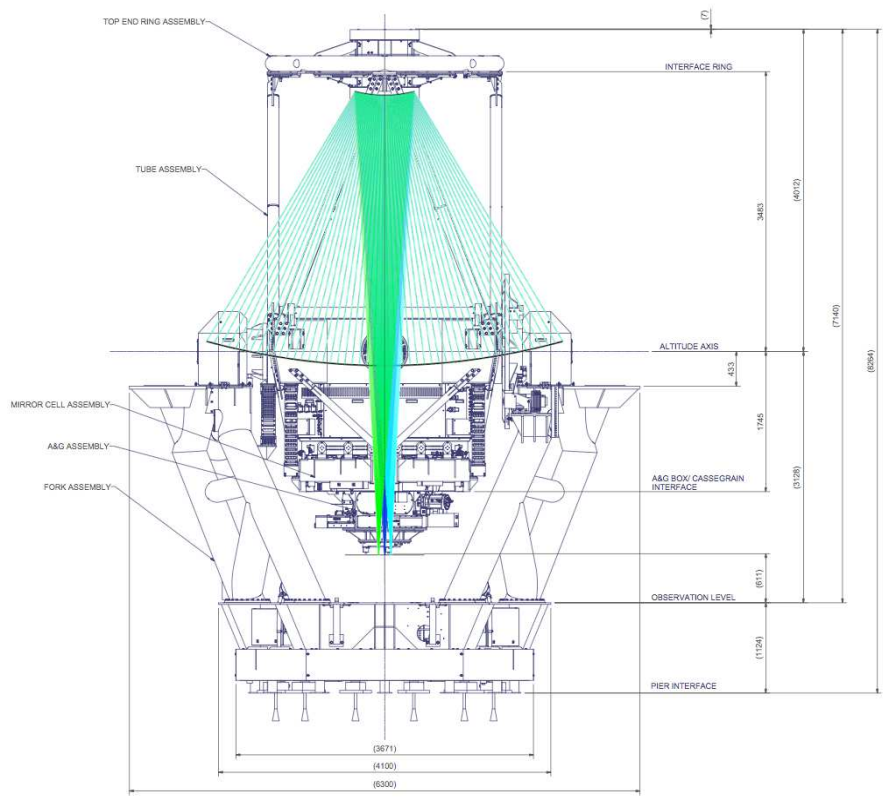}
\hfill 
\caption{The LT2 optical path, using a proposed f/8 design, superimposed on the LT at the same scale.} \label{fig:upgrade} \end{figure}

Our requirement for very fast slewing means the optical designs for LT2 are rather compact. An interesting consequence is that the space envelope required for the optics could conceivably fit within the current LT enclosure (Figure \ref{fig:upgrade}), and so alternatively we could decommission LT and reuse the existing infrastructure for LT2. A 36+ segment mirror with an effective aperture of $4$ metres is comparable in mass to the monolithic mirror of the LT, which implies the existing telescope pier could also be reused, and perhaps even some of the telescope structure itself. Replacing the LT with LT2 in the same enclosure simplifies the project significantly, and so would potentially bring delivery forward in time. However this would of course preclude any future use of the LT, and there would be a period of time ($\sim$$1$ year) in which neither telescope would be available, impacting existing scientific and educational programmes. The feasibility of this option is something we will continue to explore.

\section{Education and outreach}
\label{sec:edu}

It was decided from the outset that education and outreach would be a key part of the LT project (e.g., \citealt{Bode95}). To that end, the National Schools Observatory (NSO)\footnote{http://www.schoolsobservatory.org.uk/} was founded, and through this organisation $5$ per cent of LT observing time (recently increased to $10$ per cent) was allocated to schools in the UK and Ireland. This has been an extremely successful programme, and over the past decade thousands of schools have been provided with tens of thousands of observations. We plan to maintain our commitment to the NSO with LT2. The nature of this commitment will depend to some extent on the future of the LT (Section \ref{sec:ltfuture}). At the very least, a percentage of LT2 time will be available for schools observing. However, if the LT remains in operation then there is scope for a significant expansion of the time allocated to the NSO, since many existing science programmes will shift to the new telescope. This would potentially allow for expansion of the educational programme to schools across the European Union and beyond. A wider field-of-view LT, as proposed in Section \ref{sec:ltfuture}, would also be of general benefit to such programmes.

\section{Budget estimate}
\label{sec:budget}

We estimate the total 5-year cost for delivery of the project is $\pounds16.7$ million. Of this sum $\pounds10.4$ million is the cost of design and construction of the telescope itself. This is based on a simple top-down approach and scaling law, derived from a manufacturer's estimate of the cost to supply the primary mirror segments. We note that this cost is in good agreement with the typical cost in the literature of a 4-metre telescope \citep{Bely03}, which gives us confidence in the estimate. The remainder of the total budget incorporates staffing costs for the establishment of a project office, an instrument budget of $20$ per cent of the telescope design and construction cost, and $\pounds1.5$ million for the LT wide field upgrade. The cost of this last component of the project is based on a detailed optical design study.

\section{Summary}
\label{sec:conclusions}
We intend to lead the construction and subsequent operation and scientific exploitation of Liverpool Telescope 2: a $4$-metre class robotic facility which will build on the success of the existing $2$-metre Liverpool Telescope. We intend the telescope to be operational close to the beginning of the next decade, and it will be located at the Observatorio del Roque de Los Muchachos on the Canary island of La Palma. Robotic telescopes are ideal for time domain science, in particular rapid reaction to targets of opportunity. We are designing Liverpool Telescope 2 to be capable of extremely rapid response, typically taking data within tens of seconds of an alert. The main instrument will be an intermediate resolution, high throughput, optical/infrared spectrograph; although we will provide multiple focal stations and the capability for rapid and automatic instrument changes during the night. The telescope will be a follow-up facility for transients discovered with the next generation of optical synoptic surveys such as LSST, as well as other discovery facilities operating across the electromagnetic spectrum such as CTA, SVOM and SKA, and `multi-messenger' detectors such as aLIGO, aVirgo and IceCube.

\begin{acknowledgements}
JHK acknowledges financial support to the DAGAL network from the People Programme (Marie Curie Actions) of the European Union’s Seventh Framework Programme FP7/2007-2013/ under REA grant agreement number PITN-GA-2011-289313, and from the Spanish MINECO under grant number AYA2013-41243-P. DC is supported by a Leverhulme Emeritus Fellowship from the Leverhulme Trust. We thank the anonymous referee for their comments, which have led to a number of important improvements to this paper.
\end{acknowledgements}

\bibliography{lt2}
 
\end{document}